\documentclass[10pt,aps,a4paper,twocolumn,showpacs,superscriptaddress,floatfix,prd,nofootinbib]{revtex4-1}
\usepackage{graphicx} 
\usepackage[colorlinks=true,linktocpage=false,citecolor=blue,linkcolor=blue,urlcolor=blue,
        pdftex,             
        plainpages=false,  
        breaklinks=true,    
        pdftitle=Analysis of the proposed tests for alternative theories of gravity using LISA Pathfinder
        pdfauthor=Natalia K.
]{hyperref}
\usepackage{amssymb,amsmath}
\usepackage{booktabs,dcolumn}
\usepackage[nolist,nohyperlinks]{acronym}
\usepackage{times}
\usepackage{tikz}
\usetikzlibrary{shapes,arrows,calc}
\usepackage[normalem]{ulem}
\usepackage{units}

\newcommand{\Leps}{\text{\Large{$ \varepsilon $}}}

\newcolumntype{d}[1]{D{.}{.}{#1}} 

\newcommand{\de}[0]{\text{d}}
\newcommand{\vc}[1]{\boldsymbol{#1}}
\newcommand{\parmis}[0]{\vc{\lambda}^{\textit{m}}}
\newcommand{\partheo}[0]{\vc{\lambda}^{\textit{t}}}

\newcommand{\accunits}{\unit{m}/\unit{s}^2}
\newcommand{\kmsunits}{\unit{km}/\unit{s}} 

\everymath{\displaystyle}

\begin{document}
    
\title{Data Analysis Methods for Testing Alternative Theories of Gravity with LISA Pathfinder}

\author{Natalia Korsakova}
\email{natalia.korsakova@aei.mpg.de}
\affiliation{
Albert-Einstein-Institut, Max-Planck-Institut f{\"u}r Gravitationsphysik und Universit{\"a}t Hannover, Callinstrasse 38, 30167 Hannover, Germany
}

\author{Chris Messenger}
\affiliation{
SUPA, University of Glasgow, Glasgow, G12 8QQ, United
Kingdom
}

\author{Francesco Pannarale}
\affiliation{
School of Physics and Astronomy, Cardiff University,\\
Queens Buildings, The Parade, Cardiff, CF24 3AA, The United Kingdom
}

\author{Martin Hewitson}
\affiliation{
Albert-Einstein-Institut, Max-Planck-Institut f{\"u}r Gravitationsphysik und Universit{\"a}t Hannover, Callinstrasse 38, 30167 Hannover, Germany
}

\author{Michele Armano}
\affiliation{
ESAC, European Space Agency, Camino bajo del Castillo s/n, Urbanizaci\'on Villafranca del Castillo, Villanueva de la Ca\~nada, 28692 Madrid, Spain
}

\date{\today}

\begin{abstract}
  In this paper we present a data analysis approach
  applicable to the potential saddle-point fly-by mission extension of
  \ac{LPF}.  At the peak of its sensitivity, \ac{LPF} will sample the
  gravitational field in our Solar System with a precision of several
  $\text{fm/s}^2\text{/}\sqrt{\text{Hz}}$ at frequencies around $1\,
  \text{mHz}$. Such an accurate accelerometer will allow us to test
  alternative theories of gravity that predict deviations from
  Newtonian dynamics in the norelativistic limit. As an example, we
  consider the case of the \ac{teves} theory of gravity and
  calculate, within the nonrelativistic limit of this theory, the
  signals that anomalous tidal stresses generate in \ac{LPF}.  We
  study the parameter space of these signals and divide it into two
  subgroups, one related to the mission parameters and the other to
  the theory parameters that are determined by the gravity model. We
  investigate how the mission parameters affect the signal
  detectability concluding that these parameters can be determined
  with the sufficient precision from the navigation of the spacecraft
  and fixed during our analysis. Further, we apply Bayesian parameter
  estimation and determine the accuracy to which the gravity theory
  parameters may be inferred. We evaluate the portion of parameter
  space that may be eliminated in case of no signal detection and
  estimate the detectability of signals as a function of parameter
  space location.  We also perform a first investigation of
  non-Gaussian ``noise-glitches'' that may occur in the data. The
  analysis we develop is universal and may be applied to anomalous
  tidal stress induced signals predicted by any theory of gravity.
\end{abstract}

\acresetall 

\maketitle

\begin{acronym}
\acrodef{LPF}[LPF]{LISA Pathfinder}
\acrodef{LISA}[LISA]{Laser Interferometer Space Antenna}
\acrodef{ESA}[ESA]{European Space Agency}
\acrodef{teves}[T{\it e}V{\it e}S]{Tensor-Vector-Scalar}
\acrodef{MOND}[MOND]{MOdified Newtonian Dynamics}
\acrodef{SNR}[SNR]{signal-to-noise ratio}
\acrodef{ASD}[ASD]{amplitude spectral density}
\acrodef{PSD}[PSD]{Power Spectral Density}
\acrodef{GR}[GR]{General Relativity}
\acrodef{OMS}[OMS]{Optical Metrology System}
\acrodef{LTP}[LTP]{LISA Technology Package}
\acrodef{OSTT}[OSTT]{On-Station Thermal Tests}
\acrodef{ASD-astrium}[ASD]{Astrium Deutschland}
\acrodef{SP}[SP]{saddle point}
\end{acronym}

\section{Introduction}
\label{sec_intro}

\acf{LPF}~\cite{LPF2012} is a technology demonstration mission for
future space-based gravitational-wave observatories, such as the
\ac{LISA}. \ac{LPF} is designed to test many of the challenging
technologies needed for space-based gravitational-wave detectors and
is planned to be launched in July 2015. On the basis of the \ac{LISA}
concept, ``The Gravitational Universe'' theme (with eLISA as foreseen
implementation) was proposed to the \ac{ESA}~\cite{WhitePaper} and was
selected as a science theme for the third large-class
mission~\cite{ESAcosmicVision} to be launched in 2034 within the
\ac{ESA} Cosmic Vision science program. eLISA is a reduced version
of the original \ac{LISA} design that will nevertheless be able to
observe numerous extremely interesting sources of gravitational waves.

\ac{LPF} is a compact version of one arm of eLISA, designed to verify
the ability to place test masses in free fall at the required
sensitivity level. It consists of two equal test masses that are
accommodated within one spacecraft. The instrument measures the
relative position of two free-falling test masses with picometer
precision using laser interferometry, thus being sensitive to the
differential gradients of the gravitational potential. \ac{LPF} will
initially be placed in a Lissajous orbit around L1, the Lagrangian
point of dynamically unstable equilibrium between the Sun and the
Earth, where the gravitational forces and the centrifugal force cancel
out in the noninertial rotating reference frame.  The transition from
Earth to L1 will take three months and will be followed by six months
of experiments performed to verify the on-board technologies and
performance of the satellite~\cite{LPF_MasterPlan}. It was
noted~\cite{BM, Trenkel2009} that the combination of design solutions
for the mission, such as the sampling frequency and the overall
measurement sensitivity, would allow \ac{LPF} to probe anomalous
gravity stress tensors, i.e., ones that deviate from the Newtonian
prediction, in the low gravity regime. Anomalous stress tensors are
predicted by various alternative theories of gravity and high
precision measurements of these deviations would allow us to test such
theories. To this end, \ac{ESA} scientists and members of the science
and industrial community have been studying a possible \ac{LPF}
mission extension.  Here we consider the data analysis methods for
such a scenario.

In the solar system, the low gravity regime can be investigated at the
\acp{SP} of two-body systems, where the gradients of the gravitational
potential of two gravitating bodies are equal in magnitude and
opposite in orientation. For the Sun-Earth system, the \ac{SP} is
located about $1231000$ km away from L1 towards Earth.  A \ac{SP} is
not an equilibrium point, so it will only be possible to perform a
``fly-by'' with \ac{LPF}. When passing by the \ac{SP}, \ac{LPF} will
be sampling the gravity stress tensor in a low gravity-gradient
region. The measured variation of the distance between the two test
masses can be compared to the theoretical predictions from Newtonian
and alternative theories of gravity. From these comparisons one can
infer (i) if any deviations from Newtonian dynamics occur, and (ii)
constrain alternative theories of gravity.

The data analysis approach developed in this paper allows for a
rigorous analysis of the test made during the \ac{SP} fly-by. It aims
at exploring the possible deviations from the Newtonian dynamics by
analyzing the gravity stress tensor measured by \ac{LPF}.

We consider the class of alternative theories of gravity that have
\ac{MOND} in their nonrelativistic limit. \ac{MOND} emerged as a
possible way to explain the observations of rotational curves of
spiral
galaxies~\cite{Oort1932,Rubin1970,OstrikerPeebles,Bosma1981,Rubin1982,Zwicky2008}. The
observations show that the rotational curves of the galaxies stay
constant and do not depend on the distance from the galactic center,
as expected in Newtonian gravity. \ac{MOND} (originally proposed by
Milgrom~\cite{Milgrom1983}) is a possible heuristic solution to this
problem, in contrast to the introduction of hidden mass (i.e., dark
matter). At the core of the theory is a characteristic acceleration
$a_0 \approx 10^{-10} \accunits$ at which a transition occurs, from
the regime accurately described by the Newtonian field equation, to
one in which the gravitational dynamics is better described by a
nonlinear Poisson equation.  To embed \ac{MOND} into a consistent
theory of gravity, we chose \ac{teves} as underpinning relativistic
theory, bearing in mind that other choices could be possible. The key
details are presented in Sec.~\ref{sec_TeVeS} together with the
rationale behind our choice.

Generally speaking, alternative theories of gravity that incorporate
MOND as an additional scalar field can all be parametrized in the same
way.  In addition to the function that describes the transition from
the \ac{MOND}ian to the Newtonian regime, the contribution of the
additional scalar gravity potential introduced by these theories to
the overall physical potential will depend on two parameters. The
first parameter is also inherited from initial \ac{MOND} heuristics
and stands for the characteristic acceleration $a_0$ mentioned
earlier. The other parameter determines the coupling of the additional
scalar field to the overall physical potential. In this respect, the
analysis that is going to be performed here for the \ac{teves} theory
can be easily extended to the entire class of similar theories.

In order to study the detection of a signal of a particular shape in
additive noise, as in the \ac{LPF} \ac{SP} fly-by scenario, one must
first determine the physical quantities that influence the form of the
signal itself. In our case, we parametrize the signal in terms of two
groups of physical quantities. The first set of parameters is
determined by the way the stress tensor is sensed by the instrument
and will depend on the fly-by trajectory and the orientation of the
\ac{LPF} sensitive axis joining the two free-falling test masses. The
second set of parameters is prescribed by the theory of gravity that
determines the anomalous stress tensor under consideration and varies
from theory to theory. The parameters that come from the experiment
setup, or mission parameters, can be estimated during the flight
independently of the main scientific measurement.  The position of the
spacecraft in space as a function of time will be determined using
standard spacecraft tracking techniques, and its orientation will be
measured using on-board star trackers.  One of our goals is to
determine whether and how much the accuracy of these measurements will
influence our ability to detect a deviation from Newtonian
gravity. With this objective in mind, we quantify how mission
parameters variations will influence the measured signal and how much
this differs from the true signal, modeled using fixed values
obtained from other observations.

Primarily, we want to measure (or constrain) the second group of
parameters with \ac{LPF} and, in case of no signal detection, to draw
conclusions about the validity of a specific theory of gravity under
consideration. We chose to use a Bayesian approach to estimate the
parameter values. Further, we apply Bayes' theorem to address the
problem of model selection, in which we must choose between two
models, one that predicts the presence of a signal in the data and the
other that assumes the data to be noise only. For the analysis of the
theory parameters, the simulated data is constructed by summing
Gaussian noise, with a known amplitude spectral density, and an
anomalous tidal stress signal. We construct simulated signals by
solving the \ac{MOND} nonlinear Poisson equation [see
Eq.\,(\ref{non-Newtonian_potential})] numerically (with the help of
the code provided by our colleagues from Imperial College
London~\cite{Neil}) in a neighborhood of the Sun-Earth \ac{SP} and
by simulating the passage of \ac{LPF} along a given satellite
trajectory and with a fixed tidal stress sampling rate. We show the
parameter estimation results for several representative points in the
parameter space. We also show the outcome of the noise-only scenario
and determine the area of the parameter space that will be ruled out
in case of no signal detection. Furthermore, we present model
selection results for several points in the parameter space. Finally,
we apply the data analysis framework to realistic data from an
\ac{LPF} test campaign and discuss both parameter estimation and model
selection results. This data set is interesting as it contains a noise
artifact that can be misinterpreted by the data analysis setup as a
signal.

An important remark regarding the example of applying our data
analysis framework to the \ac{MOND} limit of \ac{teves} must be
made. Tests for alternative theories of gravity, including \ac{teves},
are performed in the strong field regime by measuring the orbital
decay of the relativistic pulsar--white dwarf binary PSR
J1738+0333~\cite{Freire2012}. The constraints imposed to the theory in
its strong field limit, however, differ from the ones that can be
imposed in the weak field limit~\cite{BlanchetBook}. The constraints
that would follow from the method described in this work would
therefore be complementary to, say, the PSR J1738+0333 ones and
largely applicable to theories exhibiting the same scalar field
coupling mechanism as \ac{teves}.

The paper is structured as follows. In Sec.~\ref{sec_lisa_pathfinder}
we discuss \ac{LPF} and explain how it performs measurements. In
Sec.~\ref{sec_parameters_lpf} we identify the mission parameters and
discuss how the trajectory of the spacecraft and the projection of the
signal on the \ac{LPF} sensitive axis will influence the signal. 
Section~\ref{subsec_data_analysis} describes the two
approaches we develop for the analysis framework of the mission and
theory parameters. 
In Sec.~\ref{sec_TeVeS}, in order to fix an example against which our
data analysis tools may be tested, we briefly describe the
non-relativistic limit of \ac{teves} theory of gravity, and we report
on the signal model construction and the space of theory parameters
for this scenario. We present our results in Sec.~\ref{sec_results}
and gather our conclusions in Sec.~\ref{sec_conclusions}, where we
also discuss possibilities of future work for this experiment.

\section{LISA Pathfinder}
\label{sec_lisa_pathfinder}

The task of measuring the residual differential acceleration of two
free-falling test masses is one of the main objectives of the \ac{LPF}
mission and, therefore, the conversion from the observed differential
displacements to differential accelerations has been analyzed in
depth~\cite{DiffAccNoise,Anneke}.

\subsection{\ac{LPF} Measurement}

\ac{LPF} measures differential displacements between two free-falling
test masses and is thus sensitive to their differential
acceleration~\cite{DiffAccNoise}. Consider the relative motion of two
masses that follow the geodesics of the gravitational field and let
the vector $\vc{\zeta}$ denote the separation between the two test
masses. The components of this vector may be expressed as
${\zeta^{i}=x^{\, i}_1-x^{\, i}_2}$, where $x^i_{\{1,2\}}$ are the
coordinates of the two test masses. Working in Cartesian coordinates,
the equations of motion for the test masses in Newtonian gravity are
\begin{equation}
  \frac{\de^2 x^{\, i}_1}{\de t^2} = -\frac{ \partial  \Phi_{\text{N}}(\vc{x}_1,t)}{\partial x^{i}}
  \label{eq_of_motion_for_TM}
\end{equation}
and
\begin{equation}
  \frac{\de^2 x^{\, i}_2}{\de t^2} = -\frac{ \partial \Phi_{\text{N}}(\vc{x}_2,t)}{\partial x^{i}}\,,
\end{equation}
where $\Phi^{\text{N}}$ is the Newtonian gravitational potential.
The relative acceleration is thus given by
\begin{equation}
\begin{split}
\label{eq:geod_dev}
\frac{\de^2 \zeta^{i}}{\de t^2}
& = \frac{\de^2 x_1^{i}}{\de t^2} - \frac{\de^2 x_2^{i}}{\de t^2} =  \\
& = -\zeta^{j} \frac{\partial^2 \Phi_{\text{N}}}{\partial x^{i} \partial x^{j}} +
o(\zeta^{i} \zeta_{i}) = -\mathcal{E}^{i}_{\: j} \zeta^{j} + o(\zeta^{i}
\zeta_{i})\,,
\end{split}
\end{equation}
where summation over repeated indices is implied, the gravitational
potential is expanded in terms of the separation vector up to the
first order, and ${ \mathcal{E}_{ij} = \partial ^2 \Phi_{\text{N}}/\partial
  x^{i} \partial x^{j}}$ is the gravitational tidal field in Cartesian
coordinates~\cite{MTW}.

\ac{LPF} has one sensitive axis that is oriented along the line
joining the two free-falling test masses. By projecting
Eq.\,(\ref{eq:geod_dev}) along this axis, one obtains
\begin{equation}
  \frac{\de^2 \zeta_{i}}{\de t^2} \hat{\zeta}^{i} = - \hat{\zeta}^{i} \zeta^{j} 
  \frac{\partial^2 \Phi_{\text{N}}(\vec{x},t)}{\partial x^{i} \partial x^{j}}\,,
\label{stress_projection}
\end{equation} 
where $\hat{\zeta}^i = \zeta^i/\| \vc{\zeta} \|$ is the $i$th component of
the unit vector in the $\vc{\zeta}$ direction.  The diagonal components
of $ \mathcal{E}_{ij}$ contribute to the relative acceleration of the
test masses, whereas the remaining components contribute to their
tilts. The diagonal components of the stress tensors are larger than
the nondiagonal ones, therefore we will consider only the relative
acceleration contribution.

\subsection{Estimation of the differential test mass acceleration}
\ac{LPF} is designed to keep the distance between the two test masses constant below $1\,\unit{mHz}$ 
by accounting for external forces, 
whereas at the sensitivity frequencies of 1--30 $\unit{mHz}$ the test masses are in free fall. 
Both test masses are accommodated within one spacecraft and free fall is
achieved by controlling the position of the spacecraft relative to one
test mass.  The position of the second test mass is then controlled
relative to the first outside the \ac{LPF} sensitive frequency band.
The differential gravitational force can thus be recovered from the
measurement of the differential displacement. An anomalous stress
tensor predicted by an alternative theory of gravity may therefore be
sensed by \ac{LPF} as the differential force acting on the test
masses.  This is performed by taking into account the
models~\cite{DiffAccNoise,Anneke} of the \ac{LPF} subsystems in the
equations of motion for the test masses along the sensitive axis,
described by
\begin{equation}
  \vc{a} = [\vc{D}^{-1}\vc{I}^{-1} + \vc{C}] \vc{o}\,,
\end{equation}
where $\vc{o} = (o_1, o_{\Delta})^{\text{T}}$ is read
interferometrically along the sensitive axis of \ac{LPF} by the two
interferometers on board, $o_1$ being the position of the first test
mass relative to the spacecraft, and $o_{\Delta}$ being the position
of the second test mass relative to the first. $\vc{a} = (a_1,
a_{\Delta})^{\text{T}}$, with $a_1 = \de^2 x_1/\de t^2$ and $a_\Delta
= \de^2 \zeta / \de t^2 $ being the estimated residual acceleration of
the spacecraft and the estimated residual differential acceleration of
the two test masses, respectively. $\vc{D}$ represents the dynamics of
the spacecraft, $\vc{I}$ the interferometer sensing matrix, and
$\vc{C}$ the controller transfer functions.  More specifically, the
dynamics of the spacecraft is \renewcommand{\arraystretch}{2.2}
\begin{equation}
  \vc{D} = 
  \left[
    \begin{matrix}
      \frac{1}{(s^2 + \omega^2_1)} & 0 \\
      -\frac{(\omega_2 - \omega_1)^2}{(s^2 + \omega_1^2)(s^2 +
        \omega_2^2)} & \frac{1}{(s^2 + \omega^2_2)}
    \end{matrix}
  \right]\,,
\end{equation}
where $s$ is a Laplace domain complex variable and ${\omega_{\{1,2\}}^2
= k_{\{1,2\}}/m}$. The mass of the test mass is $m$ and $k_{\{1,2\}}$
are the spring constants that model the gravitational and
electrostatic couplings between the test masses and the
spacecraft. Given the coupling factor $\delta$ modeling the degree to
which the differential interferometer picks up motion of the
spacecraft, the interferometer sensing matrix can be written as
\renewcommand{\arraystretch}{1}
\begin{equation}
  \vc{I} = 
  \left[
    \begin{matrix}
      1 & 0 \\
      \delta & 1
    \end{matrix}
  \right].
\end{equation}
Finally, the controller matrix that converts the measured signal into
the commanded forces may be written as
\begin{equation}
  \vc{C} = 
  \left[
    \begin{matrix}
      H_{\text{df}} & 0 \\
      0 & H_{\text{sus}}
    \end{matrix}
  \right]\,,
\end{equation}
where $H_{\text{df}}$ and $H_{\text{sus}}$ are the gains of the
drag-free and suspension control loops along the sensitive axis of
\ac{LPF}, respectively. The drag-free control loop actuates on the
spacecraft via micro-Newton thrusters, while the suspension loop
actuates on the second test mass by electrostatic actuation.

\subsection{Noise sources in the \ac{LPF} measurement}
\label{sec:2noises}
\begin{figure}
  \centering%
  \includegraphics[width=\columnwidth]{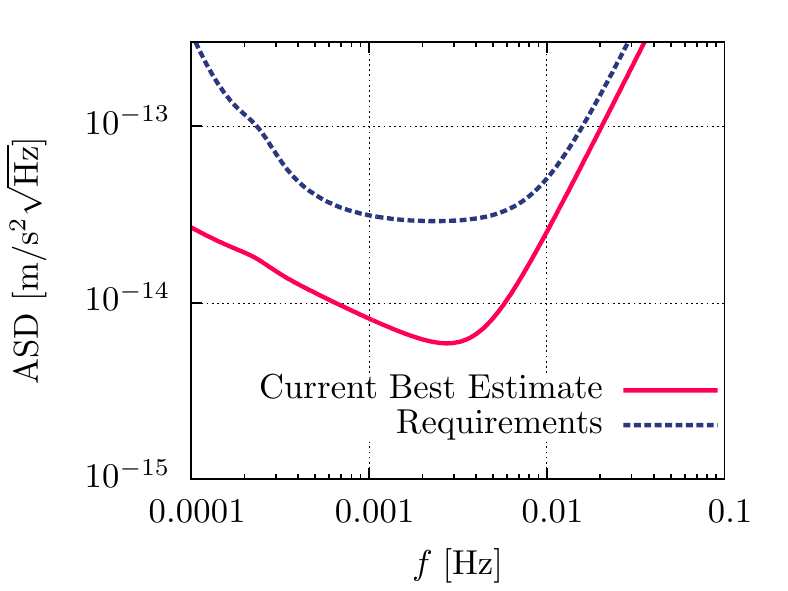}
  \caption{\ac{LPF} sensitivity. Amplitude spectral densities of the
    requirements and the current best noise
    estimates.\label{LPFnoise}}
\end{figure}

\ac{LPF} measurements are contaminated by the system noise. The design
of \ac{LPF} is such that the sensitivity of the instrument is expected
to be limited by the interferometer shot noise at high frequencies and
by force noise on the test masses at low frequencies. Various tests of
the flight hardware, however, show that the real sensitivity of
\ac{LPF} is expected to exceed the design requirements~\cite{LPF2012},
as shown in Fig.\,\ref{LPFnoise}. The noise current best estimate for
\ac{LPF} is limited by the electrostatic actuation noise on the second
test mass at low frequencies.

\section{Identification of mission parameters}
\label{sec_parameters_lpf}

\begin{figure*}
  \includegraphics[trim = 40mm 40mm 40mm 40mm,
  width=0.5\textwidth]{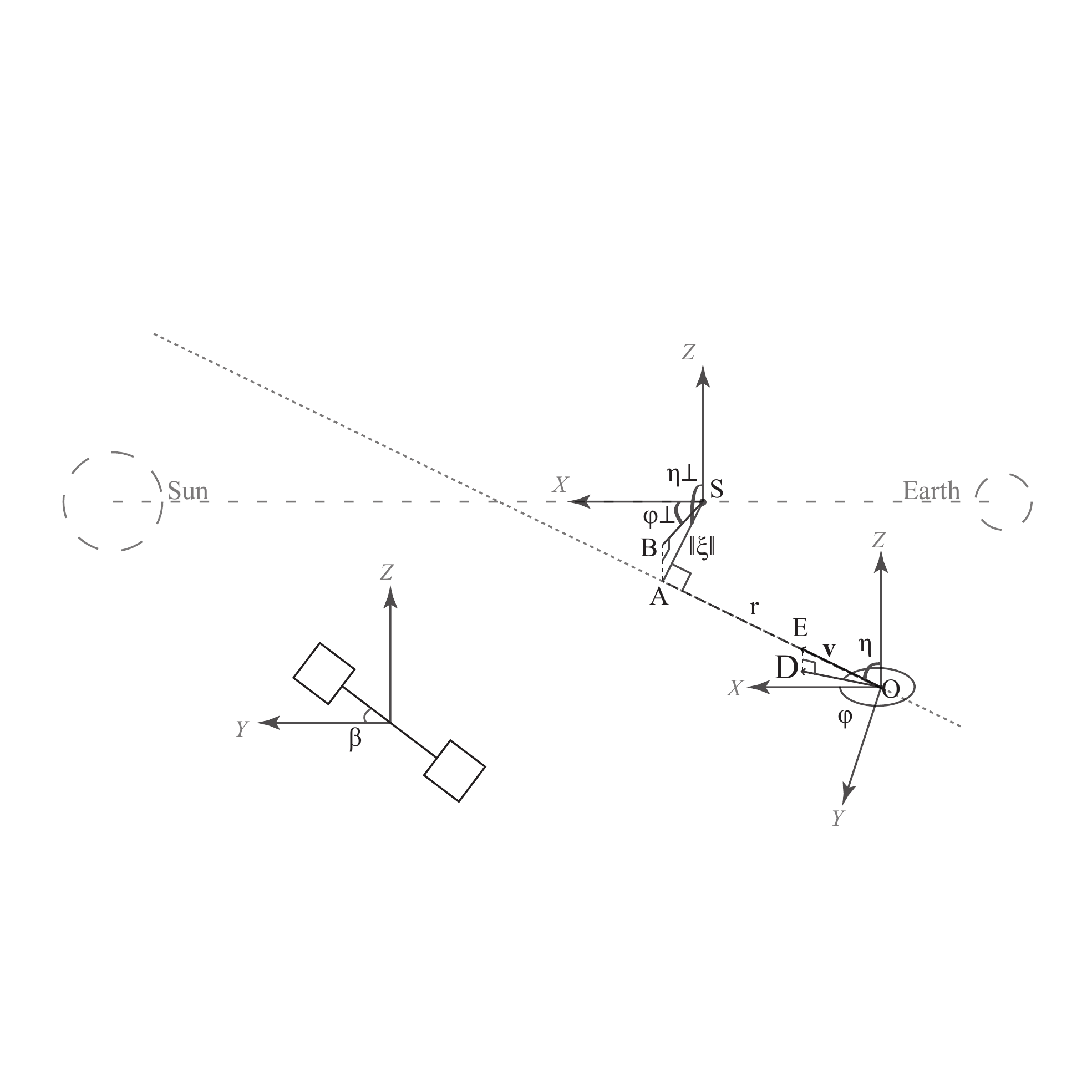}
  \caption{Schematic of the trajectory parameters. The coordinate
    system has its origin in the \ac{SP} (S) and the $x$-axis is
    parallel to the line joining the Sun and the Earth and pointing in
    the direction of the Sun. The $z$ axis is perpendicular to the
    ecliptic. The spacecraft velocity $\vc{v}$ is aligned with the
    trajectory and represented by 
      the vector ${OE}$. The direction of the trajectory is
    defined by the two angles $\eta$ (the angle between the $z$ axis
    and $\vc{v}$) and $\varphi$ [the angle between $x$ axis and
    projection of $\vc{v}$ on the $(x,y)$ plane, shown as the segment
    OD]. The position of the spacecraft along the trajectory is
    determined by the variable $r$, the distance to the point A where
    the perpendicular dropped on the trajectory intersects with
    it. The length of the perpendicular is given by the parameter
    $\|\vc{\xi}\|$. The position of the perpendicular is given by two
    angles, $\eta_\perp$ (the angle between the $z$ axis and the
    perpendicular) and $\varphi_\perp$ (the angle between the $x$ axis
    and the projection of the perpendicular on the $(x,y)$ plane,
    segment BS). The angle $\beta$ determines the projection of
      the gravity stress tensor on the sensitive axis of
      \ac{LPF}. \label{param_lpf}}
\end{figure*}

In order to parametrize the signals measured by \ac{LPF}, we must
begin by defining a method to determine the spacecraft trajectory
uniquely. Let us fix a right-handed Cartesian coordinate system with
its origin in the Sun-Earth \ac{SP}, its $x$ axis aligned with the
line connecting the Earth and the Sun, and its $z$ axis perpendicular
to the ecliptic (see Fig.~\ref{param_lpf}). The trajectory of \ac{LPF}
in the neighbourhood of the \ac{SP} can be approximated as a straight
line. The direction of the trajectory will be determined by two
angles: $\eta$, the angle between the $z$ axis and the direction of
the spacecraft velocity, and $\varphi$, the angle between the $x$ axis
and the projection of the velocity vector on the ecliptic.  The unit
vector along the trajectory of the spacecraft in the direction of
motion is, therefore
\begin{equation}
  (\hat{e}_x, \hat{e}_y, \hat{e}_z) = (\sin\eta\cos\varphi, \sin\eta\sin\varphi, \cos\eta)\,.
\end{equation}
The point of the closest approach of the trajectory to the \ac{SP},
$(\xi_x, \xi_y, \xi_z)$, determines the impact parameter, i.e.,~the
distance of the fly-by, which is the length of the perpendicular
dropped from the \ac{SP} on the trajectory. The position of the
spacecraft may thus be written as
\begin{equation}
  (x, y, z) 
  = (\xi_x, \xi_y, \xi_z) + (\hat{e}_{x}, \hat{e}_{y}, \hat{e}_{z}) r\,,
  \label{SC_trajectory}
\end{equation}
where $r$ is the distance from the point of closest approach.

Given the distance to the saddle point, the position of the closest
approach becomes redundant.  Therefore, to avoid the uncertainty the
two angles $\eta_{\perp}$ and $\varphi_{\perp}$ that define the
position of the perpendicular to the trajectory are introduced:
\begin{equation}
  \begin{split}
    & (\xi_x, \xi_y, \xi_z) = \\
    & \|\vc{\xi}\| (\sin\eta_{\perp}\cos\varphi_{\perp} ,
    \sin\eta_{\perp}\sin\varphi_{\perp} , \cos\eta_{\perp} )\,,
  \end{split}
\end{equation}
where $\|\vc{\xi}\| $ is the length of the vector $(\xi_x, \xi_y,
\xi_z)$.  Similarly to the $(\eta, \varphi)$ notation previously
introduced, $\eta_{\perp}$ denotes the angle between the perpendicular
and the ecliptic, while $\varphi_\perp$ denotes the angle between the
$x$ axis and the projection of the perpendicular on the
ecliptic. Notice that the additional condition,
\begin{equation}
  \begin{split}
    & \sin\eta\sin\eta_{\perp}(\cos\varphi\cos{\varphi_{\perp}} +
    \sin{\varphi}\sin\varphi_{\perp})+\\
    &\cos\eta\cos\eta_{\perp} = 0
    \label{perp_condition}
  \end{split}
\end{equation}
holds for the four angles $\eta$, $\varphi$, $\eta_\perp$, and
$\varphi_\perp$ as a consequence of the orthogonality between the
satellite trajectory and the line of closest approach. This allows us
to further reduce the parameters that determine the perpendicular to
the trajectory of the satellite in the neighbourhood of the \ac{SP}
down to $\eta_{\perp}$ and $\mathrm{sign}( \sin \varphi_\perp)$. The
latter determines whether $\varphi_{\perp} \in (0,\pi)$ or
$\varphi_{\perp} \in (\pi, 2\pi)$.

The signal measured by \ac{LPF} can be simulated by sampling the
stress tensor along the trajectory with velocity $v$ and the
instrument sampling frequency of $10 \; \unit{Hz}$. The velocity of
the spacecraft and the sampling frequency determine the resolution at
which the gravity stress tensor is being sampled.

As a final step, we must define the projection of the stress tensor on
the sensitive axis of \ac{LPF}. The projection is determined by the
two angles $\alpha$ and $\beta$ that the sensitive axis forms with the
$x$ axis and $y$ axis of the coordinate system, respectively. However,
since \ac{LPF} is held oriented so that its solar panel faces the Sun,
and since we are considering a neighborhood of the Sun-Earth
\ac{SP}, and because the sensitive axis of \ac{LPF} is parallel to the
solar panel, $\alpha$ can be fixed to $\alpha = 90^{\circ}$. The
projection of the stress tensor on the sensitive axis is thus
determined only by the angle $\beta$, making \ac{LPF} sensitive to the
linear combination of the two diagonal components of the stress
tensor:
\begin{equation}
  \frac{\de^2 \zeta_i}{\de t^2} \hat{\zeta}^i = -\| \vc{\zeta} \| [\mathcal{E}_{yy} \cos^2(\beta) + \mathcal{E}_{zz} \sin^2(\beta)]\,.
  \label{tens_lfpsens_proj}
\end{equation}

All in all, the signal can be fully described in terms of the
following set of mission parameters:
\begin{equation}
  \parmis_0 = \{\|\vc{\xi}\|, \eta, \varphi, \eta_{\perp}, \text{sign}(\sin \varphi_{\perp}), \|\vc{v}\|, \beta\},
  \label{lpf_parameters}
\end{equation}
which are depicted in Fig.~\ref{param_lpf}.

\section{Data Analysis}

We now introduce the approach to the analysis of the data that will be
acquired with \ac{LPF} in the vicinity of the Sun-Earth \ac{SP}.
We describe the model of the data and the derivation of a matched
filter which will be designed to study the mission
parameters. Thereafter, we develop a Bayesian approach to the analysis
of the theory parameters.

\subsection{Data model}\label{sec:data_model}
\label{subsec_data_analysis}

The detector noise is modeled as having a frequency dependent
spectrum (see Fig.~\ref{LPFnoise}), hence it is more natural to carry
out the analysis in the frequency domain.  We write the measured data
as
\begin{equation}
  \tilde{x}(f,\vc{\lambda}_0) = \tilde{h}(f,\parmis_0,\partheo_0) + \tilde{n}(f)\,,
  \label{signal_model_freq}
\end{equation}
where $\tilde{h}(f,\parmis_0,\partheo_0)$ and $\tilde{n}(f)$ are the
Fourier transforms of the signal and the detector noise, respectively.
$\vec{\lambda}_{0}=(\parmis_0,\partheo_0)$, where $\parmis_0$ and
$\partheo_0$ denote the mission and the theory parameters that govern
the signal: the former are listed in Eq.\,(\ref{lpf_parameters}),
whereas the latter will be discussed in the course of the paper.  We
model the noise as Gaussian, with zero mean and two-sided noise power
spectral density
\begin{equation}
  S(f)\approx \vert\tilde{n}(f)\vert^2 /\Delta f\,,
  \label{variance2PSD}
\end{equation}
where $\Delta f = 1/T$ is the size of the frequency bin, whereas $T =
N \cdot \Delta t$ with $N$ -- the number of samples over the measurement time
interval $[0,T]$ and $\Delta t$ -- the time domain sampling interval. The Fourier transform of the noise averaged over ensemble is the variance of the noise $\sigma^2 = < \vert\tilde{n}(f)\vert^2 >$. The noise models we use are defined by the
theoretical \ac{ASD} shown in Fig.~\ref{LPFnoise}.

In order to test our data analysis framework on artificial data, we
must choose a model to produce signal templates.  As anticipated in
the Introduction, in this paper we consider the stress tensor
predictions obtained within the nonrelativistic limit of Bekenstein's
\ac{teves} theory of gravity. This theory embeds the heuristic
description of the dynamics of galaxies provided by MOND into a
consistent relativistic theory (see Appendix~\ref{appendix_teves}).

\subsection{Building signal templates}
\label{sec_TeVeS}

\subsubsection{Nonrelativistic limit of \ac{teves}}

As we are going to perform the experiment in the Solar System, we must
consider the quasistatic, weak potential, and slow motion limit of
\ac{teves}~\cite{TeVeS}. We may thus take the metric to be time
independent. Additionally, as we work in a neighborhood of the Sun-Earth \ac{SP}, 
far enough from both bodies, we may set the metric to
be flat. In the nonrelativistic limit, the full physical potential
that determines the test particle acceleration within \ac{teves}, $
\vec{a} = -\vc{\nabla}\Phi$, is given by the sum of the Newtonian
vector potential $ \Phi_\text{N} $ and the scalar potential $ \phi$,
i.e.,
\begin{equation}
  \Phi = \Phi_\text{N} + \phi + \mathcal{O}(\Phi^2_\text{N})\,.
  \label{Phi_PhiN_phi}
\end{equation}
Therefore \ac{LPF} will be measuring $\Phi$, which has to replace $\Phi_{\text{N}}$ in Eqs.(~\ref{eq_of_motion_for_TM})--(\ref{stress_projection}).
The Newtonian potential is given by the familiar Poisson equation
\begin{equation}
  \vc{\nabla}^2 \Phi_\text{N} = 4\pi G\tilde{\rho}\,,
  \label{Poisson_equation}
\end{equation}
where $ \tilde{\rho} $ is the baryonic mass density, whereas the
scalar potential $ \phi $ is determined by the nonlinear Poisson
equation
\begin{equation}
  \vc{\nabla}\cdot \left[\mu\left(k l^2 (\vc{\nabla} \phi)^2\right)\vc{\nabla} \phi\right] = k G \tilde{\rho}\,,
  \label{non-Newtonian_potential}
\end{equation}
where $ k $ is a dimensionless constant and $ l $ is a constant
length.

The $\mu$ function appearing in the last equation is a free function
that governs the transition from the Newtonian regime to the
\ac{MOND}ian one [see Eq.\,(\ref{phi_eom})].  We can reparametrise its
dimensionless argument $ y \equiv kl^2 (\vc{\nabla} \phi)^2 $ in terms
of an acceleration parameter
\begin{equation}
  a_0 \equiv \frac{(3k)^{1/2}}{4\pi l}\,,
  \label{a_0}
\end{equation}
thus obtaining
\begin{equation}
  \label{eq:2TheoryParameters}
  y = 3\left(\frac{k}{4\pi}\right)^2\left( \frac{\vc{\nabla}\phi}{a_0}\right)^2\,,
\end{equation}
where the ratio between the \ac{MOND}ian acceleration and the
acceleration parameter is now manifest. The asymptotical limits of the
$\mu$ function must therefore obey the following requirements:
\begin{equation}
    \begin{array}{ll}
      \mu(y) \rightarrow  1, & \; \text{for} \; y \rightarrow \infty, \\
      \mu(y) \approx \sqrt{y/3}, & \; \text{for} \; y \ll 1,
    \end{array}
    \,
  \label{mu_conditions}
\end{equation}
where the first condition leads to the Newtonian regime.  The second
condition ensures that in the low acceleration regime, i.e. $
|\vc{\nabla} \Phi| \ll a_0$, the \ac{MOND} modification originally
proposed by Milgrom \cite{Milgrom1983} generates a different dynamics,
recovering, for example, the one exhibited by rotational curves of
galaxies.

\subsubsection{Signal Model and Parameter Space}
\label{subsection_sm_and_ps}

\begin{figure}[!t]
  \centering \includegraphics[width=\columnwidth]{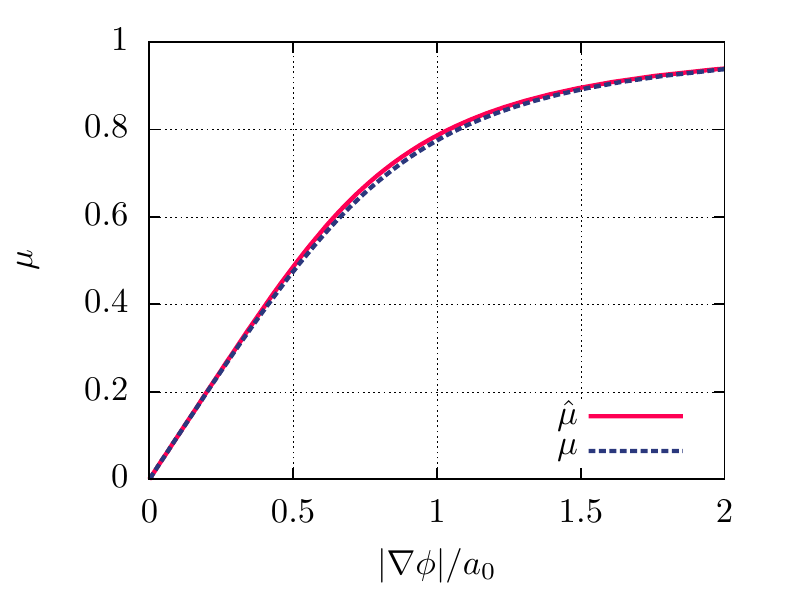}
  \caption{Comparison between the interpolating function used for the
    numerical calculations and the one originally proposed
    in~\cite{TeVeS}.}
  \label{mu_functions_comparison}
\end{figure}

As shown by
Eqs.\,(\ref{non-Newtonian_potential})--(\ref{eq:2TheoryParameters}),
within the example selected for this paper, the signal models will be
determined by two parameters $ k $ and $ a_0$, and a free function, $
\mu $. For the moment, we fix the $ \mu$ function to the form that was
proposed in~\cite{TeVeS}. In terms of the notation introduced in
Eq.\,(\ref{signal_model_freq}), therefore, $ \partheo_0=\{k,a_0\}
$. The nonlinear elliptical differential equation which determines
the scalar potential $ \phi $ and hence the tidal stress tensor,
Eq.\,(\ref{non-Newtonian_potential}), can be solved
numerically~\cite{Neil} (the code that implements the numerical
solution was kindly provided by Imperial College London). While
in~\cite{TeVeS} the $ \mu$-function definition is
\begin{equation}
  y = \frac{3}{4}\frac{\mu^2 (\mu - 2)^2}{1 - \mu}\,,
  \label{mu_TeVeS}
\end{equation}
the interpolating function $ \mu $ in our numerical calculations is
fixed via the relation
\begin{equation}
  \frac{\hat{\mu}}{\sqrt{1-\hat{\mu}^4}} = \frac{k}{4\pi} \frac{ \vert \vc{\nabla}\phi\vert}{a_0}\,,
  \label{mu-func_num}
\end{equation}
where we used the notation $ \hat{\mu} $ to explicitly distinguish
this function from the one appearing in Eq.\,(\ref{mu_TeVeS}). As
shown in Fig.~\ref{mu_functions_comparison}, the two functions are in
a good agreement. The advantage of $ \hat{\mu} $ is that it may be
written out analytically as
\begin{equation}
  \hat{\mu} = \sqrt{\frac{-1+\sqrt{1+4x^2}}{2x}}\,,
\end{equation}
where $ x=y/3$. In solving the nonlinear Poisson equation
numerically, the condition $ \mu = \sqrt{x} $ for $ x < 10^{-5} $ is
used [see Eq.\,(\ref{mu_conditions})].

To solve Eq.\,(\ref{non-Newtonian_potential}) numerically, other than
fixing the $ \mu$ function, we must prescribe boundary conditions.  We
use the rescaled Newtonian potential for this purpose. This is readily
obtained from
Eqs.\,(\ref{Poisson_equation}) and (\ref{non-Newtonian_potential}) by
taking into account that $ \mu \rightarrow 1 $ as $
|\vc{\nabla}\phi|/a_o\rightarrow\infty $ and by applying Gauss's
theorem. This yields
\begin{equation}
  \vc{\nabla}\phi = \frac{k}{4\pi}\vc{\nabla}\Phi_\text{N}\,,
  \label{phi_PhiN}
\end{equation}
so that the gradient of the physical potential $ \Phi $ reduces to the
usual Newtonian form with a renormalised gravitational constant given
by
\begin{equation}
  G_\text{N} = \left(1 + \frac{k}{4\pi}\right)G\,.
  \label{bigG_rescaling}
\end{equation}

In order to produce signal templates for \ac{LPF}, as a first step we
compute the spatial derivatives of $ \vc{\nabla}\phi $ at each grid
point. This provides the nine stress tensor components, namely,
$ \partial^2 \phi / \partial x_i \partial x_j$, where $ x_{i,j} =
{x,y,z}$, at each point of the lattice. Once this is done, we must
prescribe values for the set of mission parameters listed in
Eq.\,(\ref{lpf_parameters}) and sample the stress tensor along the
\ac{LPF} trajectory [Eq.\,(\ref{SC_trajectory})]. The sampling points
are determined by the the spacings $ \|(\Delta x, \Delta y, \Delta
z)\| = \| \vc{v} \|\Delta t$, with time step $ \Delta t =
1/f_{\text{samp}}$, $ f_{\text{samp}} = 10\,\unit{Hz} $ being the
\ac{LPF} sampling frequency. The stress tensor components are
calculated at each sampling point by performing a trilinear
interpolation on a three-dimensional irregular grid. The interpolation
procedure starts with a linear interpolation in the $ x$ axis
direction. This is followed by a linear interpolation along the $
y$ axis employing the $ x$-interpolated values. Finally, both the $
x$- and $ y$-interpolated values are used to perform the linear
interpolation in the $ z $ direction.

Our goals are (1) to see how the signal templates change when varying
the two theory parameters $ k $ and $ a_0$, and (2) to study their
detectability in the noise. The value of the dimensionless coupling
constant $ k $ should be of the order $ 10^{-2} $ to be consistent
with the cosmological expansion; $ k=0.03 $ is chosen
in~\cite{TeVeS}. The characteristic acceleration is usually set to $
a_0 \approx 10^{-10} \; \accunits$, in accordance with observations of
rotational curves of galaxies~\cite{StacyMcGaugh}. We vary both
parameters within reasonable ranges around their ``original'' values,
so that $ k \in [0; 0.12] $ and $ a_0 \in [0; 4\cdot 10^{-10}] \;
\accunits$. We cover this two dimensional space of theory parameters
with a $ 9\times 9 $ uniform grid (see
Fig.~\ref{figure_grid_and_restrictions}) and solve
Eq.\,(\ref{non-Newtonian_potential}) numerically in the neighbourhood
of the Sun-Earth \ac{SP} for all choices of $
(k,a_0)$.\footnote{Calculations were performed using~\cite{Cluster}.}
We then fix a set of trajectory parameters and produce \ac{LPF} signal
templates by projecting the computed stress tensor as in
Eq.\,(\ref{tens_lfpsens_proj}), at all points in the $ (k,a_0) $
parameter space. Additionally, we set $ \partial \phi^2/\partial
x_i \partial x_j = 0 $ along $ k = 0 $ and $ a_0 = 0\; \accunits$, as
proposed in~\cite{TeVeS}.  In order to obtain signal templates for
generic values of $ k $ and $ a_0$, we use a bicubic interpolation
along both directions. We interpolate the signal templates from the
knows solutions for the stress tensor on the two-dimensional parameter
space. The interpolation is performed for each sample in the template
time series. This is possible since, for a given set of trajectory
parameters, a sample in the template time series represents the same
position in time and in space for a particular choice of $ a_0 $ and $
k$.

\begin{figure}
  \centering \resizebox{\columnwidth}{!}{%
    \includegraphics[width=\columnwidth]{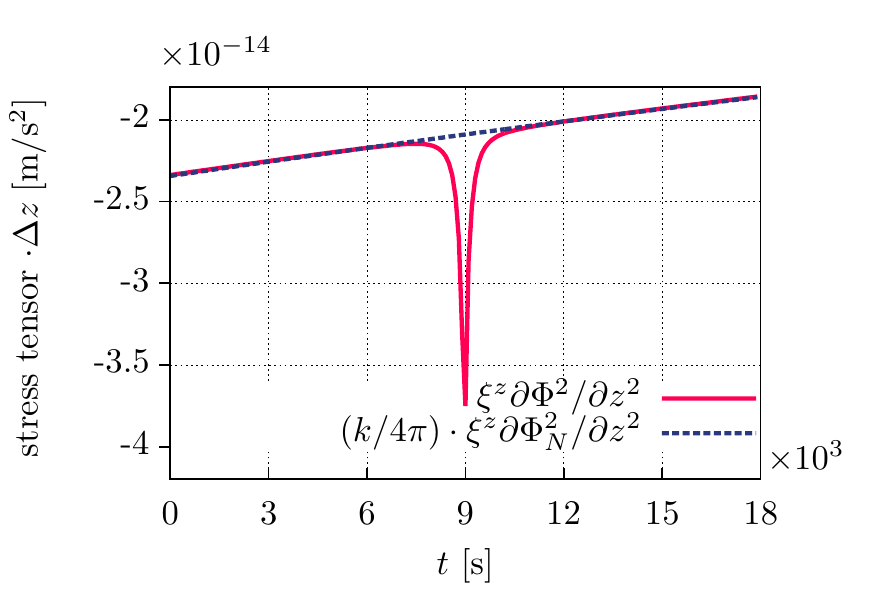}}
  \caption{Comparison between a template produced with a numerical
    calculation and the rescaled Newtonian background analytically
    estimated using Eq.\,(\ref{Newtonian_st}). In this example, $ k =
    0.03 $ and $ a_0 = 10^{-10} \; \accunits$. The $ \partial
    \Phi^2/ \partial z^2 $ and $ \partial \Phi_N^2/\partial z^2 $
    components of the \ac{MOND}ian and Newtonian stress tensors are
    plotted. This means that the sensitive axis is parallel to the $
    z$ axis of the coordinate system and, therefore, that $ \beta =
    0$.}
  \label{template_and_newtonian_st}
\end{figure}

\begin{table*}[!t]
  \caption{This table lists the seven mission parameters also shown graphically in Fig.\,\ref{param_lpf} and provides estimates for their uncertainties, the ranges in which their values are varied to produce Fig.~\ref{snr_trajectory_together}, and the values assigned to them during our parameter estimation analyses. These parameters can be determined from measurements of the spacecraft position which are based on the spacecraft navigation system without involving the \ac{LPF} optical readout~\cite{Christian,Christian_Stephen}. The uncertainties on the navigation parameter values before the flight, i.e. before the trajectory for the transition from L1 to \ac{SP} is chosen, and those determined during the flight are provided in columns three and four, respectively. The errors on the angle $\alpha$ that defines the orientation of the solar panel are below $1^\circ$: as explained in Sec.~\ref{sec_parameters_lpf}, we set $\alpha=90^\circ$ and the error may be neglected within the scope of this paper. Additionally, the time of closest approach to the \ac{SP} is not included in the parameter list as it is of the order of several seconds and can be neglected with respect to the signal length. The values reported in the last column are those used for the analysis of the theory parameters. These numbers are based on~\cite{Trenkel2012} and~\cite{Christian}. While, recent investigations show that it may be possible to realize a trajectory directly through the \ac{SP}, we have conservatively set $ \| \xi \| = 20\,\unit{km} $. \label{table_lpf_mission_parameters}}
  \resizebox{\textwidth}{!}{%
    \begin{tabular}{c@{\hspace{0.25cm}}c@{\hspace{0.25cm}}c@{\hspace{0.25cm}}c@{\hspace{0.25cm}}c@{\hspace{0.25cm}}c}
      \toprule[1.pt]
      \toprule[1.pt]
      \addlinespace[0.3em]
      Parameter & Description & Uncertainty before flight & Uncertainty after flight & Range & Value \\
      \addlinespace[0.2em]
      \midrule[1.pt]
      \addlinespace[0.2em]
      $ \|\vc{\xi}\| $                    & Fly-by distance                                 & $ 5\,\unit{km} $ & $ 5\,\unit{km} $ & $ [0;300]\,\unit{km} $  & $ 20\,\unit{km} $ \\
      $ \varphi $                    & Trajectory polar angle                          & $ 30^{\circ} $ & $ \ll 1^{\circ} $ & $ [0;360]^{\circ} $          & $ 30^{\circ} $ \\
      $ \eta $                       & Trajectory azimuthal angle                      & $ 30^\circ $ & $ \ll 1^\circ $ & $ [0;180]^{\circ} $          & $ 70^{\circ} $ \\
      $ \eta_\perp $                  & Polar angle of the position of closest approach & uniform     & $ \| \xi \| $ dependent & $ [0;180]^{\circ} $          & $ 90^{\circ} $ \\
      $ \text{sign} (\sin \varphi_\perp ) $ & Hemisphere of the position of closest approach  & $ \{-1,1\} $ & \dots & $ \{-1,1\} $                & $ +1 $ \\
      $ \| \vc{v} \| $                    & Spacecraft velocity                             & $ 0.1\,\kmsunits $ & $ 1\,\unit{cm}/\unit{s}$   & $ [1;2]\,\kmsunits $ & $ 1.5\,\kmsunits $ \\
      $ \beta $                      & Orientation of the \ac{LPF} sensitive axis      & $ 30^\prime $ & $ 30^\prime $ & $ [0;360]^{\circ} $          & $ 0^{\circ} $ \\
      \bottomrule[1.pt]
      \bottomrule[1.pt]
    \end{tabular}
  }
\end{table*}

As a final remark, we note that in some instances the choice of the
theory parameters requires to extend the templates outside the lattice
where the \ac{MOND}ian stress tensor is calculated. As this extension
must be performed in a Newtonian limit regime, we exploit the scaling
relation between the Newtonian stress tensor (analytically computed,
see Appendix~\ref{appendix_Newtonian_ST}) and the \ac{MOND}dian one:
these are related by a factor $ k/4\pi $ [see
Eqs.\,(\ref{phi_PhiN})-(\ref{bigG_rescaling})], so that projecting the
rescaled Newtonian stress tensor along the \ac{LPF} sensitive axis
allows us to extend the \ac{MOND}ian template. An example of this is
shown in Fig.~\ref{template_and_newtonian_st}.

\subsection{Analysis of the mission parameters}
\label{subsec_mission_param_analysis}

In this section we study how the template of the predicted signal
changes when varying the mission parameters. This knowledge will
validate our choice in studying the theory and the mission parameters
independently. This greatly simplifies the study of theories that
predict signals that can be measured with \ac{LPF}. To investigate the
mission parameter space we fix the theory parameters to $k = 0.03$ and
$a_0 = 10^{-10}\,$m/s$^2$, following~\cite{TeVeS}. In this section,
for the sake of simplicity, we also remove references to the theory
parameters from the notation.

We begin by introducing the concept of a linear filter. In terms of
our problem, it is a signal template with a certain set of parameters.
Its construction is based on the ``true'' signal that has a fixed set
of (mission) parameters $\parmis_0$. In order to quantitatively assess
the influence of parameter variations, we estimate the response of the
filter to ``data'' generated using mission parameters
$\parmis_{\text{v}}$ that have an offset $\Delta \parmis =
\parmis_{\text{v}} -\parmis_0$ within the range of spacecraft
navigation errors reported in Table
\ref{table_lpf_mission_parameters}. This table provides the accuracy
with which each parameter can be determined from navigation system
measurements. We report both the errors on the mission parameters
assigned before the flight ({\it Uncertainty before the flight}) and
the precision attainable during the flight by spacecraft navigation
system measurements ({\it Uncertainty after the flight})
\cite{Christian,Christian_Stephen}. Notice that the low precision on
the angles $\varphi$ and $\eta$ before the flight follows from the
uncertainty on the trajectory which depends on the departure
conditions from the Lissajous orbit around L1~\cite{Christian} and
they will be known better once the trajectory is chosen.

The correlation between the data, $\tilde{x}$, and a signal template,
$\tilde{q}$, can be calculated as the output of a matched filter via
\begin{equation}
  C(\tau, \Delta \parmis) = \int_{-\infty}^{\infty} \tilde{x}(f,  \parmis_{\text{v}})\tilde{q}^{*}(f,  \parmis_0)e^{-2\pi if\tau}\de f.
\end{equation}
The signal at the output of the matched filter is the averaged
correlation function, for which $\langle
\tilde{x}(f,\parmis_{\text{v}}) \rangle = \langle
\tilde{h}(f, \parmis_{\text{v}})\rangle $ since $\langle \tilde{n}(f)
\rangle = 0$.  We do not take into account the time delay $\tau$ of
the signal arrival. We assume that the expected time of the signal
arrival, which is the time when the spacecraft has its closest
approach to the \ac{SP} is known.  The error on the time of the signal
arrival is embedded in the parameter that defines the distance from
the \ac{SP} to the point where the measurement is made.  The mean of
the correlation function between the data on the output of the
instrument and the linear filter $\tilde{q}$~\cite{Sathya1993} thus
reads
\begin{equation}
  \hat{C}(\Delta \parmis) = \int_{-\infty}^{\infty} \tilde{h}(f,\parmis_{\text{v}}) \tilde{q}^*(f,\parmis_0) \de f.
\end{equation}

By setting the linear filter to the true template weighted by the
noise power spectral density, i.e.
\begin{equation}
  \tilde{q}^{*}(f,\parmis_0) = \frac{\tilde{h}^{*}(f,\parmis_0)}{S(f)}\,,
\end{equation}
the filter becomes optimal~\cite{Sathya_Bernard}.  An optimal matched
filter is one that maximizes the \ac{SNR}
\begin{equation}
  \rho^2 = \hat{C}(\Delta \parmis = 0) =  \int_{-\infty}^{\infty}  \frac{ \tilde{h}(f,\parmis_0) \tilde{h}^{*}(f,\parmis_0)}{S(f)}\de f\,.
  \label{SNR_def}
\end{equation}

In the case of optimal filtering, one searches for the filter that
best fits the data. This provides a way to estimate the true
signal template. In our study, fixing the true signal template a
priori and building a filter upon it allows us to determine the
dependency of the magnitude of the matched filter response to a signal
with its parameters offset by $\Delta
\parmis$. This is the measure generally used to
quantify the resolution with which we can distinguish one template
from another. With this in mind, we rewrite the filter in discrete
form,
\begin{equation}
  c(\Delta\parmis) = c(\parmis_0, \parmis_{\text{v}}) = \sum_{j = 1}^{N} \frac{ \tilde{h}(f_j,\parmis_{\text{v}}) \tilde{h}^*(f_j,\parmis_0) }{  S(f_j)  } \Delta f_j\,,
\end{equation}
where frequency indices cover the instrument frequency range and
$\Delta f_j = f_{j+1} - f_j$, and we consider the ambiguity function
built upon the linear filter as follows:
\begin{equation}
  \hat{c}(\parmis_0,\parmis_{\text{v}})  = 
  \frac{ c(\parmis_0,\parmis_{\text{v}})}
  {\sqrt{  c(\parmis_0,\parmis_0)  c(\parmis_{\text{v}},\parmis_{\text{v}})}}\,.
  \label{ambiguity_function}
\end{equation} 
The ambiguity function is normalized to yield unity when the template
matches the input signal and less than unity otherwise.

\begin{figure}[!b]
  \resizebox{\columnwidth}{!}{%
    \includegraphics[width=\columnwidth]{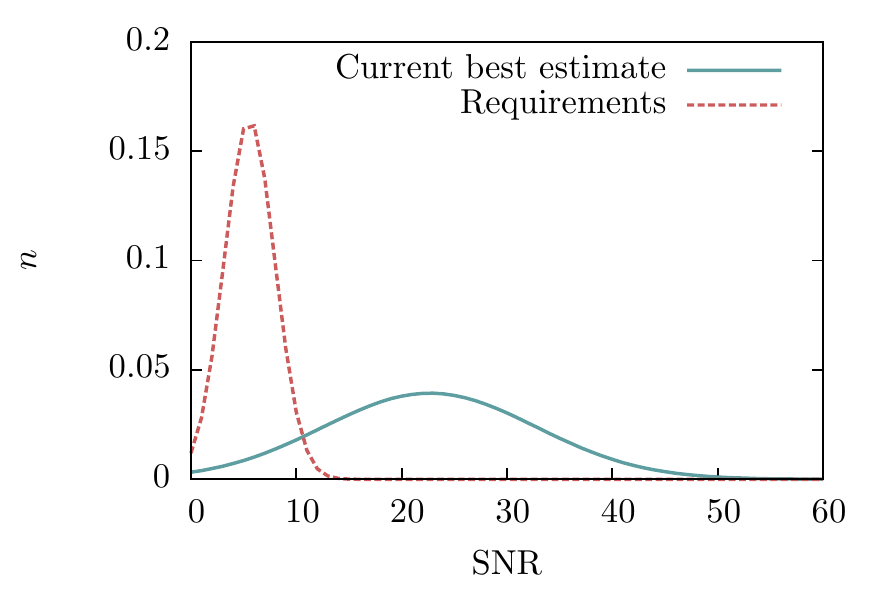}}
  \caption{Fraction of trajectories $n$ with \ac{SNR} value specified
    on the horizontal axis.  The \ac{SNR} was calculated using $1000$
    trajectories with randomly varied parameters for both the current
    best noise estimate and the requirements noise. The parameter
    values were uniformly sampled over the ranges given in the fifth
    column of Table \ref{table_lpf_mission_parameters}. The curves are
    the Gaussian fits to the discrete distributions that were
    obtained.}
  \label{snr_trajectory_together}
\end{figure}

\begin{figure}[!b]
  \resizebox{\columnwidth}{!}{%
    \includegraphics[width=\columnwidth]{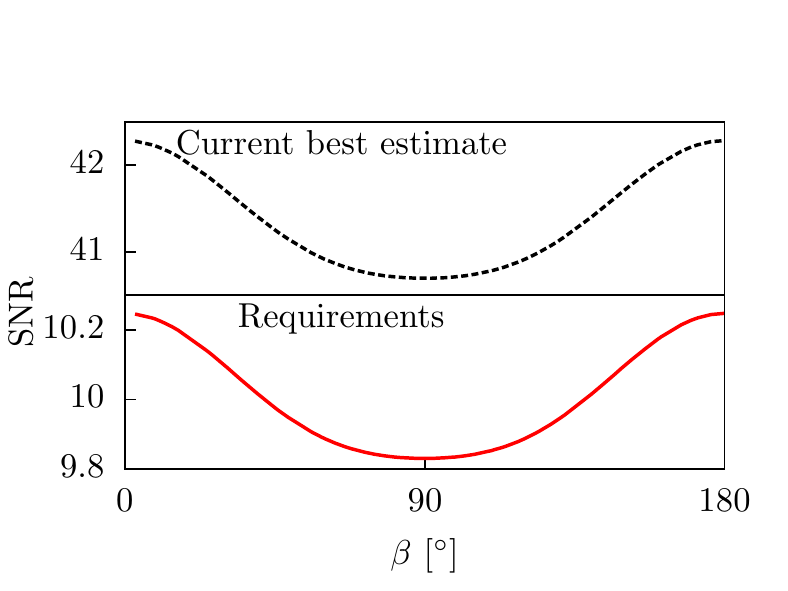}}
  \caption{\ac{SNR} as a function of the orientation angle of the
    sensitive axis $\beta$ for the two noise realizations. The
    remaining mission parameters are fixed according to the set of
    values given in the last column of Table
    \ref{table_lpf_mission_parameters}.}
  \label{snr_beta}
\end{figure}

\subsubsection{SNR as a function of mission parameters}
Estimating the \ac{SNR} as a function of the mission parameters
provides insight into the optimal values these should take and allows
us to identify any peculiar behavior of the templates over the
parameter space. In turn, if no peculiarities emerge, we assume that
this allows us to investigate the behavior of the signal in the
neighborhood of a single, representative point of our choice in the
parameter space and to extrapolate results over the whole range of
parameter values. At this location of our choice, we investigate the
behavior of the ambiguity function, as this allows us to assess how
much reduction in \ac{SNR} would be caused by deviations from the
nominal mission parameter values.

We now compute the expected \ac{SNR} for the two noise models --
current best noise estimate and requirements noise -- discussed in
Sec.~\ref{sec:2noises}.  The \ac{SNR} values are calculated using
Eq.\,(\ref{SNR_def}) for $1000$ different trajectories each with
random parameter values uniformly sampled within the ranges given in
the fifth column of Table \ref{table_lpf_mission_parameters}.  As
shown in Fig.~\ref{snr_trajectory_together}, the Gaussian fits to the
histograms of the \ac{SNR} values peak at $\rho \simeq 23$ and $\rho
\simeq 5$ for the current best noise estimate and the requirements
estimate, respectively.

When varying the mission parameters sequentially within the predefined
ranges, the remaining parameters are fixed to the values given in the
last column of Table \ref{table_lpf_mission_parameters}.

\begin{figure}
  \resizebox{\columnwidth}{!}{%
    \includegraphics[width=\columnwidth]{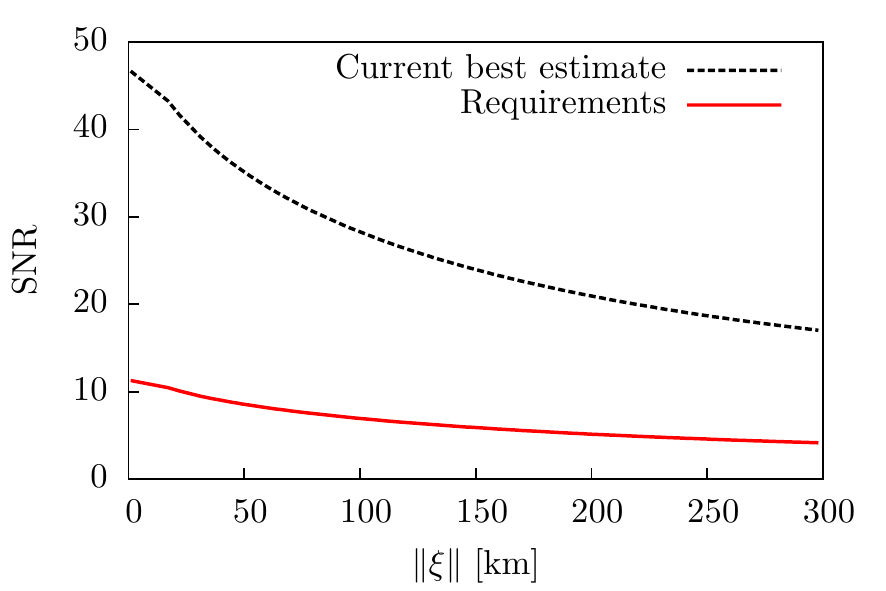}}
  \caption{\ac{SNR} as a function of the distance from the \ac{SP}
    $\|\vc{\xi}\|$ for the two noise realizations. The remaining
    parameters are fixed according to the set of values given in the
    last column of Table \ref{table_lpf_mission_parameters}.}
  \label{snr_d}
\end{figure}

\begin{figure}
  \resizebox{\columnwidth}{!}{%
    \includegraphics[width=\columnwidth]{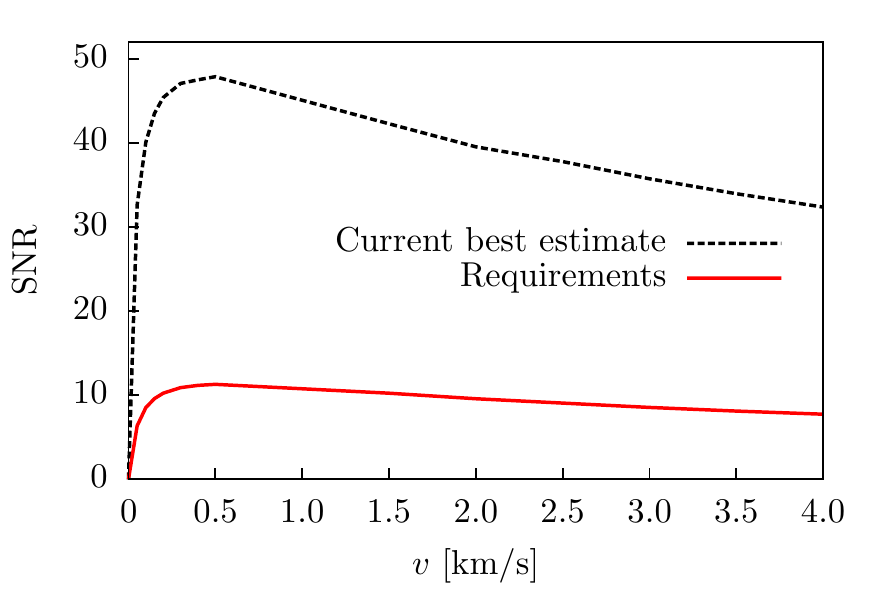}}
  \caption{\ac{SNR} as a function of the spacecraft velocity $v$ for
    the two noise realizations. The remaining parameters are fixed
    according to the set of values given in the last column of Table~
    \ref{table_lpf_mission_parameters}.  We vary the values of
    velocity within the larger range than given in the
    Table~\ref{table_lpf_mission_parameters}, i.e. from $0$ to $4$
    km/s, to observe the maximum of \ac{SNR}.}
  \label{snr_v_larger_range}
\end{figure}

The first parameter we vary is the sensitive axis orientation angle
$\beta$. As seen in Fig.~\ref{snr_beta}, the \ac{SNR} is not very
sensitive to the choice of $\beta$ and that the optimal value for
$\beta$ for both noise realisations is $ \beta = 0^{\circ} $ or $\beta
= 180^{\circ}$. We will thus fix $\beta=0^\circ$ for the analysis and
for the experiment planning.

The \ac{SNR} exhibits a smooth behaviour also when the fly-by distance
and the spacecraft velocity are varied, as shown in Figs.~\ref{snr_d}
and~\ref{snr_v_larger_range}, respectively. We notice that, as is to
be expected, the closer \ac{LPF} flies to the \ac{SP}, the higher the
\ac{SNR} is, because tidal stress deviations are stronger, whereas the
specific value of the spacecraft velocity is not very crucial in the
interval reported in Table \ref{table_lpf_mission_parameters}.

\begin{figure}
  \resizebox{\columnwidth}{!}{%
    \includegraphics[width=\columnwidth]{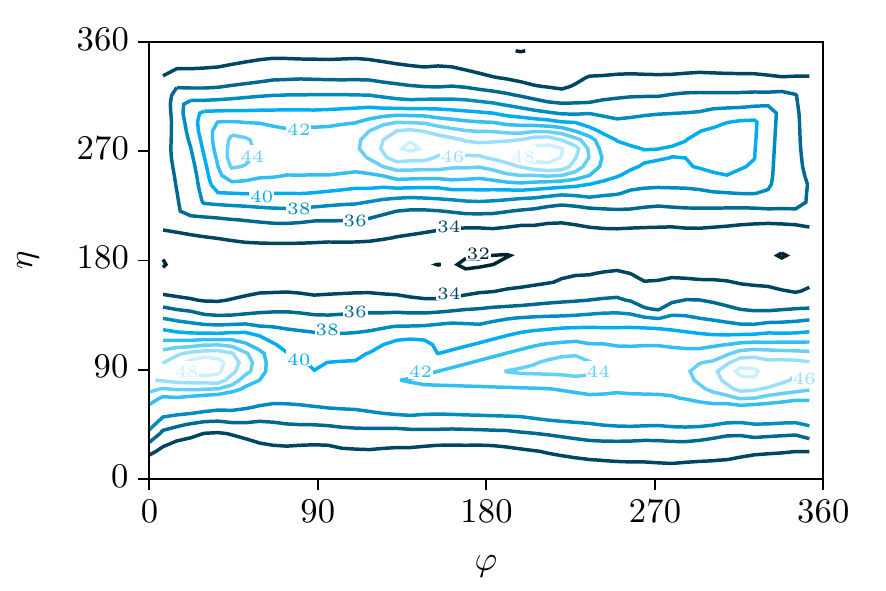}}
  \caption{\ac{SNR} as a function of the angles $\varphi$ and $\eta$
    that determine the direction of the trajectory. The \ac{SNR}
    estimates are plotted for the current best noise estimate. The
    behaviour for the requirements noise is similar, but with
    magnitudes in the range $[8;12]$. The remaining parameters are
    fixed to the set of values given in the last column of Table
    \ref{table_lpf_mission_parameters}.}
  \label{contour_phieta}
\end{figure}

Similarly, the \ac{SNR} is smooth in the $\varphi$-$\eta$ subspace, as
shown in Fig.~\ref{contour_phieta}. These are the two angles that
define the orientation of the spacecraft trajectory. While the
\ac{SNR} is flat in $\varphi$, it is maximum for
$\eta=\{90^\circ,270^\circ\}$.  In these specific cases we see that
more \ac{SNR} is accumulated if \ac{LPF} flies within the Ecliptic
plane and that the direction of flight within this plane has minimal
influence.

As the range of values covered by $\eta_{\perp}$ depends on the
combination of other parameter values via Eq.\,(\ref{perp_condition}),
$\eta_{\perp}$ cannot span the whole interval $[0,180]^\circ$ for a
specific choice of $\eta$ and $\varphi$. Therefore, we do not present
\ac{SNR} estimates as a function of $\eta_{\perp}$. We note, however,
that in the cases we considered the dependence of the \ac{SNR} on
$\eta_{\perp}$ is weak.

\subsubsection{\ac{SNR} loss due to mismatched mission parameters}

Having established the dependence of the \ac{SNR} on the mission
parameter space, we may now study the loss of \ac{SNR} as a function
of parameter mismatch within the known navigation uncertainties on the
mission parameters. As discussed previously, we fix $\beta =
0^{\circ}$. At the same time, even though $\eta$ has its highest
\ac{SNR} estimate for $\eta = 90^{\circ}$, we will choose it to be
$\eta = 70^{\circ}$ in order to avoid performing our analyses in the
best case scenario. Contrary to the alignment of the \ac{LPF}
sensitive axis, the value of $\eta$ depends on the manouvres that are
necessary for \ac{LPF} to leave the Lissajous orbit around the first
Lagrangian point. Further, the option of multiple
fly-by's~\cite{Christian} implies different estimates for the angle
values.  We therefore keep this parameter away from its optimal value
during our analyses and avoid choosing a trajectory within the
Ecliptic plane.

\begin{figure}
  \resizebox{\columnwidth}{!}{%
    \includegraphics[width=\columnwidth]{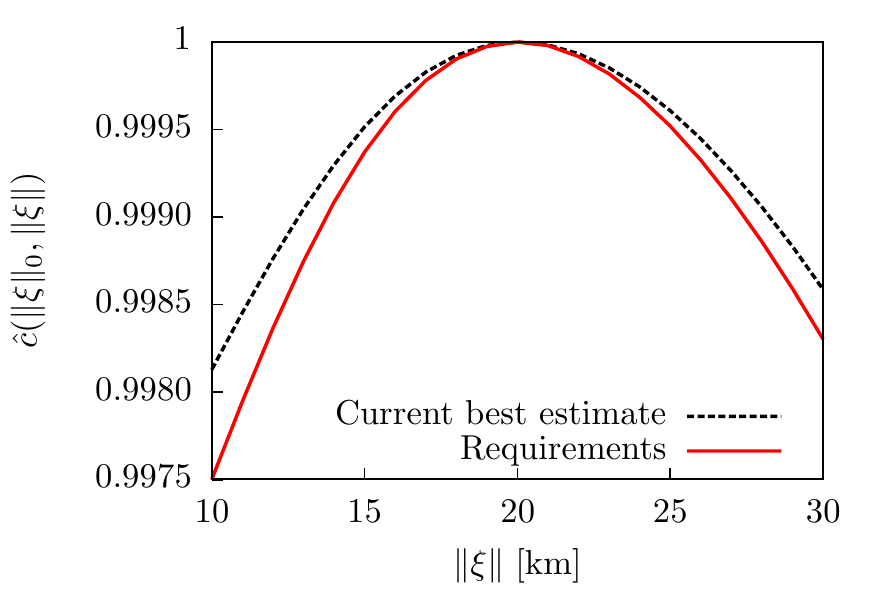}}
  \caption{Ambiguity function for the fly-by distance $\|\vc{\xi}\|$
    for the two noise realizations. The true value of the parameter is
    $\|\vc{\xi}\|_0 = 20\,\unit{km}$. The remaining parameters are
    fixed according to the set of values given in the last column of
    Table \ref{table_lpf_mission_parameters}.}
  \label{ambig_d}
\end{figure}

Hereafter, we proceed by taking one-dimensional slices through the
parameter space, fixing six parameters out of seven to the values
listed in the last column of Table
\ref{table_lpf_mission_parameters}. The parameters are varied only
around their true values, i.e.~the values listed in Table
\ref{table_lpf_mission_parameters}, which we treat as the parameters of
the signal buried in the data. All parameters are varied within
intervals that include the spacecraft navigation errors listed in
Table \ref{table_lpf_mission_parameters}. Similarly to what we did for
\acp{SNR}, we estimate the ambiguity function
[Eq.\,(\ref{ambiguity_function})] between templates with varied
parameter values and the template with all parameters set to the
values listed in Table \ref{table_lpf_mission_parameters}. When the
ambiguity function varies very little, we can assume the parameters
are essentially exactly known and can be fixed during the analysis of
the theory parameters.

\begin{figure}[!t]
  \resizebox{\columnwidth}{!}{%
    \includegraphics[width=\columnwidth]{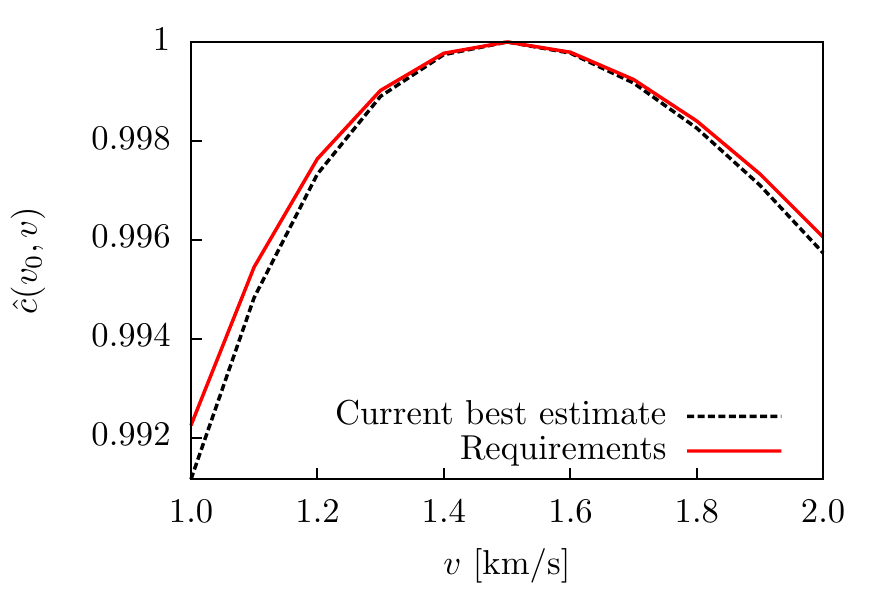}}
  \caption{Ambiguity function for the spacecraft velocity $v$ for the
    two noise realizations. The true value of the parameter is $v_0 =
    1.5\,\kmsunits$. The remaining parameters are fixed according to
    the set of values given in the last column of Table
    \ref{table_lpf_mission_parameters}.}
  \label{ambig_v}
\end{figure}
 
\begin{figure}
  \resizebox{\columnwidth}{!}{%
    \includegraphics[width=\columnwidth]{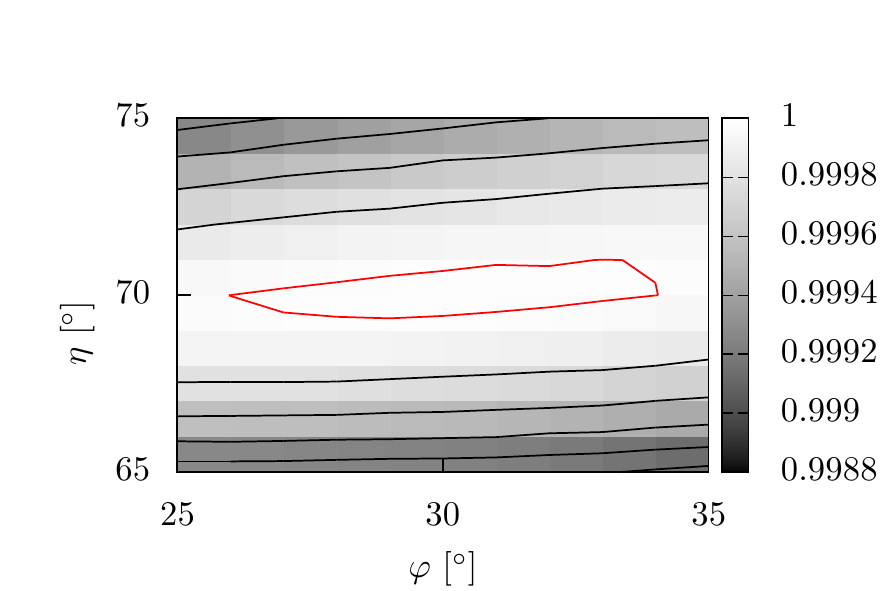}}
  \caption{Two-dimensional ambiguity function for the angles $\varphi$
    and $\eta$ that determine the direction of the spacecraft
    trajectory. The results are obtained with the current best noise
    estimate model. The true values of the parameters are set to
    $\varphi_0 = 30^{\circ}$ and $\eta_0 = 70^{\circ}$ and the
    remaining parameters are fixed according to the set of values
    given in the last column of Table
    \ref{table_lpf_mission_parameters}. Both angles are varied with
    steps of $1^{\circ}$. The closed contour indicates the location of
    $\hat{c} = 0.99998$.}
  \label{ambig_phieta}
\end{figure}

Our results for the fly-by distance $\|\vc{\xi}\|$ are shown in
Fig.~\ref{ambig_d}. The true values of the mission parameters follow
Table \ref{table_lpf_mission_parameters}, so that $\|\vc{\xi}\|_0 =
20\,\unit{km}$. Templates were evaluated between $\|\vc{\xi}\|=
10\,\unit{km}$ and $\| \vc{\xi} \|=30\,\unit{km}$ every $1\,\unit{km}$
and the ambiguity function $\hat{c}(\| \vc{\xi} \|_0, \|\vc{\xi}\|)$
was calculated correspondingly, using both \ac{LPF} noise curves. We
find that if the fly-by distance is mismatched by less then
$5\,\unit{km}$, i.e.,~the navigation error before the flight reported
in Table \ref{table_lpf_mission_parameters}, the ambiguity function is
greater than $0.999$. We conclude that we can fix this parameter to
$20\,\unit{km}$ for future analyses and that it does not need to be
estimated from the \ac{LPF} measurement, but can instead be determined
via the spacecraft navigation system.
 
The same conclusion holds for the spacecraft velocity $v$. We set $v_0
= 1.5\,\kmsunits$ to be the true value of the parameter and calculate
the ambiguity function $\hat{c}(v_0,v)$ varying $v$ between
$1.0\,\kmsunits$ and $2\,\kmsunits$ and sampling it every
$0.1\,\kmsunits$. The results are shown in Fig.~\ref{ambig_v} for both
\ac{LPF} noise realizations. As is evident, templates are more
sensitive to velocity uncertainties and variations. However,
$\hat{c}(v_0,v) > 0.998$ for velocity variations within
$0.1\,\kmsunits$, which is the value reported in Table
\ref{table_lpf_mission_parameters} for the uncertainty before the
flight. Further, $v$ may be determined during the flight with an
uncertainty of $1\,$cm/s, so we assume this parameter to be fixed at
$1.5\,\kmsunits$ during future analyses.
 
Next, we vary the angles $\varphi$ and $\eta$ that determine the
orientation of the spacecraft trajectory. Our results for the
ambiguity function are presented in Fig.~\ref{ambig_phieta}. The true
parameter values are $\varphi_0 = 30^{\circ}$ and $\eta_0 =
70^{\circ}$. We consider an interval of $10^\circ$ around both values
and sample each interval every $1^\circ$. The contours shown in the
figure are for the current best noise estimate. The elongation
relative to the ecliptic changes the template more than the angle the
defines the inclination to the line connecting the Earth and the
Sun. Despite the big uncertainty in these parameters before the
experiment (see Table \ref{table_lpf_mission_parameters}), the errors
on the determination of these parameters during flight are very small
($\ll 1^{\circ}$), so that they, too, may be assumed to be fixed to
their true values for future analyses. The result for the requirements
noise is very similar to the result for the current best noise
estimate, therefore we will not display them here.
  
\begin{figure}[!t]
  \resizebox{\columnwidth}{!}{%
    \includegraphics[width=\columnwidth]{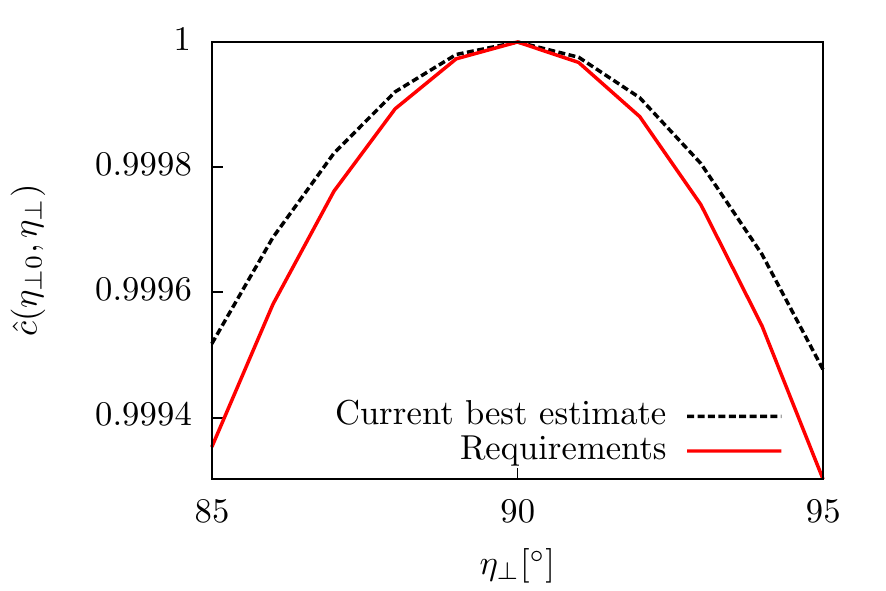}}
  \caption{Ambiguity function for the angle $\eta_{\perp}$ which
    defines the position of the perpendicular to the trajectory for
    the two noise realizations. The true value of the parameter is
    $\eta_{\perp} = 90^{\circ}$ The remaining parameters are fixed
    according to the set of values given in the last column of Table
    \ref{table_lpf_mission_parameters}.}
  \label{ambig_eta_p}
\end{figure}

Finally we consider the position of the perpendicular to the
trajectory determined by $\text{sign}(\sin\varphi_\perp)$ and
$\eta_\perp$.  For $\text{sign}(\sin\varphi_\perp)$ there will be no
uncertainty after the flight and for the $\eta_\perp$ the results are
presented in Fig.~\ref{ambig_eta_p}.  They show that the signal
templates are not sensitive to variations of this angle.
 
To summarize, we picked a specific location in the mission and theory
parameter space and investigated the behavior of the ambiguity
function.  Within the predicted uncertainties on the mission
parameters reported in Table~\ref{table_lpf_mission_parameters}, the
ambiguity function drops minimally compared to the case of exactly
matching templates. By assuming that this is the case for all other
possible parameters space locations, we make the accurate
approximation that the mission parameters can be assumed to be
``known'' without any loss of generality.  They are no longer search
parameters, which leaves only the theory parameters as unknowns and as
the sole target of the search. The analysis of the theory parameters
will therefore not require the mission parameters to be measured, nor
will it need them to be considered during parameter estimation and
model selection. In other words, we can factor the mission parameters
out of the theory parameter analyses.

Additionally, we were able to determine the optimal values of $\beta$
-- the \ac{LPF} sensitive axis orientation -- and $\eta$ -- the angle
between the spacecraft trajectory and the perpendicular to the
Ecliptic plane. In the latter case, we showed that the optimal
trajectory lies in the plane of the cliptic.

\subsection{Analysis of the theory parameters}
\label{subsec_theory_param_analysis}

We now discuss the data analysis framework to study the signal
predicted by various alternative theories of gravity. We apply this
framework to the case of the \ac{teves} theory. More specifically,
having fixed an interpolating function $\mu$, we study the $(k,a_0)$
parameter space, where $k$ is a dimensionless coupling parameter and
$a_0$ is a characteristic acceleration scale (see
Sec.~\ref{subsection_sm_and_ps}). We introduce a parameter estimation
method based on a Bayesian approach.  With this method, information
regarding the parameters of the theory can be extracted from the
data. Further, we exploit Bayes' theorem to perform model selection,
choosing between the hypothesis of having a signal in the noise and
the null hypothesis according to which the data consists of noise
only.

We discuss how parameter estimation results can be assessed in the
case of absence of a signal and how this allows us to rule out
portions of the parameter space. Finally, we show how model selection
can be applied to realistic data that contains noise artifacts. The
results of this study will show whether a glitch in the data can be
misinterpreted as a signal and where this will be localized in the
parameter space.

\subsubsection{Bayesian parameter estimation}

Following Bayes' theorem, the posterior distribution
$p(k,a_0|\{\tilde{x}\},I)$ of $k$ and $a_0$ given the data $\{
\tilde{x}\}$ and the relevant background information $I$ reads
\begin{equation}
  p(k,a_0|\{\tilde{x}\},I) = 
  \frac{p(\{\tilde{x}\}|k,a_0,I)p(k,a_0|I)}{p(\{ \tilde{x}\}|I)}\,,
  \label{posterior}
\end{equation}
where $p(k,a_0|I)$ is the prior distribution on the parameters,
$p(\{\tilde{x}\}|k,a_0,I)$ is the likelihood, and $p(\{
\tilde{x}\}|I)$ is the Bayesian evidence, which is the marginal
probability density of the data and normalizes the posterior. The data
model is the sum of a deterministic signal and Gaussian noise and is
computed in the frequency domain, as described in
Sec.~\ref{sec:data_model}. We therefore write the likelihood of the
Fourier transformed data $\{ \tilde{x}\}$ as
\begin{equation}
  \begin{split}
    p(\{\tilde{x}\}|k,a_0,I) &= \\
    \prod_{j=1}^{N/2}\frac{1}{\sigma_j^22\pi} &\exp \left(
      -\frac{\left|\tilde{x}_j -
          \tilde{h}_j(k,a_0)\right|^2}{2\sigma_j^2} \right)\,,
  \end{split}
\end{equation}
where $N$ is the number of samples over the measurement time interval.
In this expression, the variance of the noise $\sigma_j^2$ is
calculated from the \ac{PSD} normalized by the width of the frequency
bin ${\sigma^2_j = S(f_j)/\Delta f}$ [see
Eq.\,(\ref{variance2PSD})]. The noise model is based on the
theoretical estimates of the noise for \ac{LPF} (see
Fig.~\ref{LPFnoise}).  In writing the expression for the likelihood,
we assumed that each frequency bin is statistically independent, so
that the likelihood can be written as the product of bivariate
Gaussian probability density functions.

As a result of the parameter estimation, we shall obtain a joint
posterior distribution for parameters $k$ and $a_0$. However, we are
also interested in estimating each parameter separately after
performing the experiment. To obtain the posterior distribution of
each parameter separately, we marginalize the joint distribution for
the two parameters over the other parameter, i.e.
\begin{subequations}
  \begin{align}
    p(k|\{\tilde{x}\},I) &= \int_{-\infty}^{\infty} p(k,a_0|\{\tilde{x}\},I)\de a_0 \label{k_distribution}\\
    p(a_0|\{\tilde{x}\},I) &= \int_{-\infty}^{\infty}
    p(k,a_0|\{\tilde{x}\},I)\de k \label{a0_distribution}\,.
  \end{align}
\end{subequations}
These marginal distributions represent our belief in a specific value
of one of the two parameters and yield the uncertainty on the
parameter estimate following the experiment.

\subsubsection{Prior space}
As a first step to set priors in the ($k,a_0$) parameter space, we
restrict it using the following considerations. We assume that, within
some precision, the gradient of the gravitational potential is
Newtonian in the nonrelativistic limit at a distance from the \ac{SP}
equal to the distance from the \ac{SP} to the Earth.  The gradient of
the non-Newtonian potential at this distance depends on the parameters
$k$ and $a_0$ and allows us, therefore, to impose restrictions on the
combination of these parameters. Eq.\,(\ref{non-Newtonian_potential}),
which governs the non-Newtonian potential $\phi$, depends on the
$\mu$ function, which goes to unity in the Newtonian limit, when its
argument becomes sufficiently large. Taking the definition\footnote{We
  remark that the interpolating function used in the numerical
  calculations defined in Eq.\,(\ref{mu-func_num}) and the one
  expanded here correspond in the limit we consider, as shown in
  Fig.~\ref{mu_functions_comparison}.}  of the interpolating function
$\mu$ given in Eq.\,(\ref{mu_TeVeS}) and expanding it in the $\vert
\vc{\nabla}\Phi\vert/a_0 \gg 1$ limit, when $\mu \rightarrow 1$, we
obtain
\begin{equation}
  y = \frac{3}{4(1 - \mu)} + \mathcal{O}(1 - \mu)\,.
\end{equation}
Equations (\ref{Phi_PhiN_phi}) and (\ref{phi_PhiN}) can then be used to
express the argument of the $\mu$ function as
\begin{equation}
  y \equiv k l^2 \vert \vc{\nabla} \phi\vert^2 = \frac{k^3l^2}{16\pi^2} \vert \vc{\nabla} \Phi \vert^2,
\end{equation}
where higher order corrections in $(k/4\pi)$ are neglected. Combing
the last two results and expressing $l$ in terms of $a_0$ as in
Eq.\,(\ref{a_0}) yield
\begin{equation}
  \mu \approx 1 - \frac{64 \pi^4}{k^4} \frac{a_0^2}{\vert \vc{\nabla} \Phi\vert^2} + \mathcal{O}\left(y^{-2}\right)\,.
\end{equation}
If we fix an admissible error $\Leps^2$ on deviations of $\mu$ from
unity, we readily obtain the constraint
\begin{equation}
  \frac{a_0}{\vert \vc{\nabla} \Phi\vert} < \frac{k^2}{8 \pi^2} \Leps\,.
  \label{equation_boundary}
\end{equation}
Imposing this restriction allows one to exclude certain combinations
of $k$ and $a_0$.

\begin{figure}
  \centering \resizebox{\columnwidth}{!}{
    \includegraphics[width=\columnwidth]{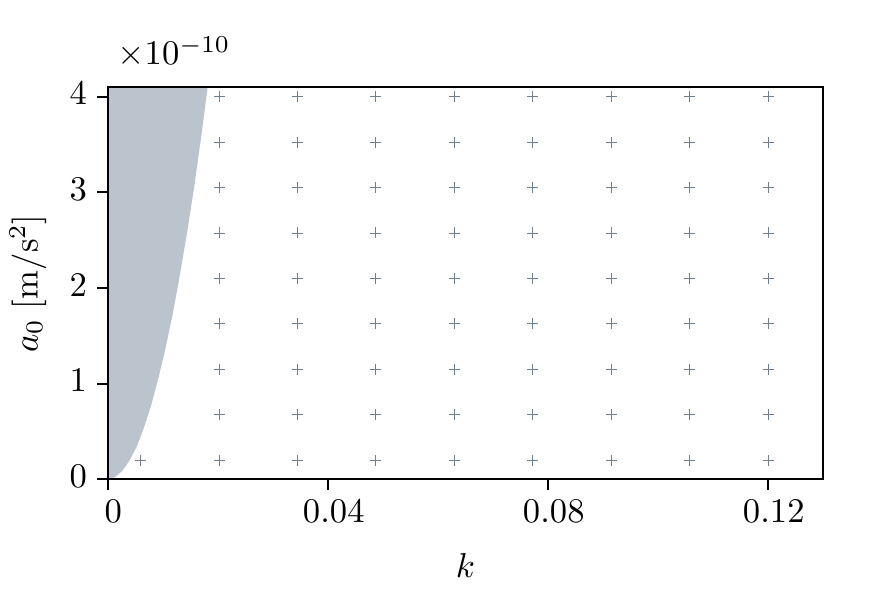}}
  \caption{The $(k, a_0)$ parameter space. The shaded area represents
    the part of the parameter space ruled out by
    Eq.\,(\ref{equation_boundary}). Crosses indicate points where
    Eq.\,(\ref{non-Newtonian_potential}) was solved
    numerically. Signal templates are built upon these solutions and
    are used, in turn, to determine signal templates at a generic
    point ($k, a_0$) via bicubic interpolation.}
  \label{figure_grid_and_restrictions}
\end{figure}

In our analysis, we set $\Leps = 10^{-5}$, and the resulting,
restricted parameter space is shown in
Fig.~\ref{figure_grid_and_restrictions}. This is a conservative value
compared to the latest boundaries imposed on the precision of the
additional acceleration allowed in the Solar
System~\cite{Fienga2010}. We do not take into account such stringent
requirements, as we want to develop and illustrate a data analysis
scheme that does not automatically depend on other astronomical
restrictions of the parameter space.

We consider a uniform prior parameter distribution (known as flat or
constant prior) for the theory parameters. We thus set the prior for
$a_0$ and $k$ to be flat in the admissible portion of the parameter
space $\mathcal{P}$, the area of which is given by
\begin{equation}
  \mathcal{A}= \int_{0}^{k^{\text{min}}} a^\prime_0(k)  \text{d}k + (k^{\text{max}} - k^{\text{min}})(a_{0}^{\text{max}} - a_{0}^{\text{min}}),
  \label{prior_area}
\end{equation}
where $k^{\text{min}}$ is the value for which
{$a^\prime_0(k^{\text{min}}) = a_0^\text{max} = 4\times
  10^{-10}\accunits$} and $a^\prime_0(k)$ is a solution of
Eq.\,(\ref{equation_boundary}).  Moreover, the values of the modified
stress tensor are set at the lower boundary of the parameter space
$a_0^{\text{min}} = 0$ to $\partial^2 \phi / \partial x^i \partial x^j
= 0$.  It reflects the \ac{GR} limit of \ac{teves} that can be
obtained when $l \rightarrow
\infty$~\cite{TeVeS}. Eq.\,(\ref{a_0}) shows that this
corresponds to $a_0 \rightarrow 0$.  We therefore have
\begin{equation}
  p(k,a_{0}| I) =
  \begin{cases}
    1/ \mathcal{A} &  (k,a_{0})\in\mathcal{P}\\
    0 , & \text{otherwise}.
  \end{cases}
\end{equation}

Flat priors depend on no underlying knowledge on the parameters,
except the assumptions made on their span. As discussed in
Sec.~\ref{subsection_sm_and_ps}, the ranges for the theory parameters
is chosen here on the basis of astrophysical
observations~\cite{StacyMcGaugh} and in order to keep the theory
consistent~\cite{TeVeS}.

As we consider a constant prior, with the exception of the prior
boundary constraints, the shape of the posterior parameter
distributions will be dictated only by the likelihood function. We
note that our Bayesian analysis scheme allows for more physically
realistic priors which opens a way for the future analyses of
different theoretical models.

\subsection{Model selection}
\label{model_selection}

The framework for model selection that we develop here is based on the
Bayesian approach to model selection and can be applied to a variety
of hypotheses. For example, we can test a model that assumes the data
is the sum of a signal and Gaussian noise, a model that assumes that
the data is Gaussian noise only, a model that assumes the data is
non-Gaussian noise, a model that assumes Gaussian noise with glitches,
and so forth.

Any number of models $\mathcal{M}_i$ can be defined and Bayes' theorem
[see Eq.\,(\ref{posterior})] can be directly applied as follows:
\begin{equation}
  p(\mathcal{M}_i\vert \{\tilde{x}\},I) 
  = \frac{p(\{\tilde{x}\} \vert \mathcal{M}_i, I)p(\mathcal{M}_i \vert I)}{p(\{\tilde{x}\} \vert I)}\,.
  \label{Bayes_theorem_with_models}
\end{equation}
This expression tells us how to determine the posterior probability
$p(\mathcal{M}_i\vert \{\tilde{x}\},I)$, which is the probability of
the $i$th model $\mathcal{M}_i$ being correct, given the data
$\{\tilde{x}\}$ and the background information $I$. The denominator is
the Bayesian evidence, a normalization term that reads
\begin{equation}
  p(\{\tilde{x}\}\vert I) 
  = \sum_{i}p(\{\tilde{x}\} \vert \mathcal{M}_i, I)p(\mathcal{M}_i \vert I)\,,
  \label{Bayesian_ev_model_selection}
\end{equation}
where $p(\{\tilde{x}\} \vert \mathcal{M}_i, I)$ is the evidence for
the model $\mathcal{M}_i$ and $p(\mathcal{M}_i \vert I)$ is the model
prior.

To properly normalise the model posterior distribution, however, one
must know all possible models in order to compute
Eq.\,(\ref{Bayesian_ev_model_selection}) and hence
Eq.\,(\ref{Bayes_theorem_with_models}). This may be avoided by
considering the ratio between model posteriors, usually referred to as
posterior odds ratio. For two models $\mathcal{M}_1$ and
$\mathcal{M}_2$, this reads
\begin{equation}
  \frac{p(\mathcal{M}_1\vert\{\tilde{x}\},I)}{p(\mathcal{M}_2 \vert \{\tilde{x}\},I)} = 
  \frac{p(\{\tilde{x}\}\vert\mathcal{M}_1,I)}{p(\{\tilde{x}\}\vert \mathcal{M}_2,I)}
  \frac{p(\mathcal{M}_1\vert I)}{p(\mathcal{M}_2\vert I)}\,.
  \label{model_selection_ratio_odds}
\end{equation}
The ratio between the evidences for the two models appearing on the
right hand side of the equation is called the Bayes factor. The second
fraction on the same side of the equation, $p(\mathcal{M}_1\vert
I)/p(\mathcal{M}_2\vert I)$, is the prior model odds.  The posterior
odds ratio represents our confidence in one model against the other,
based on the data and the background information $I$. Here
$p(\{\tilde{x}\}\vert\mathcal{M},I)$ is the likelihood marginalized
over its entire parameter space for each model.

As our goal is to quantify our confidence in signal detection, we
introduce two ways to model the measured data. The first model,
labeled $\mathcal{S}$, describes the data as the sum of a signal and
of Gaussian noise, i.e.,
\begin{equation}
  \tilde{x}_j = \tilde{h}_j + \tilde{n}_j\,.
\end{equation}
The second model, with label $\mathcal{N}$, describes the data as
Gaussian noise only, that is,
\begin{equation}
  \tilde{x}_j = \tilde{n}_j\,.
\end{equation}
The ratio between the $\mathcal{S}$ and $\mathcal{N}$ model posteriors
is thus
\begin{equation}
  \frac{p(\mathcal{S}\vert\{\tilde{x}\},I)}{p(\mathcal{N} \vert \{\tilde{x}\},I)} = 
  \frac{p(\{\tilde{x}\}\vert\mathcal{S},I)}{p(\{\tilde{x}\}\vert \mathcal{N},I)}
  \frac{p(\mathcal{S}\vert I)}{p(\mathcal{N}\vert I)}\,.
  \label{model_selection_ratio_bayes}
\end{equation}

The Bayesian evidence for a model is calculated by integrating the
joint probability density for the data and parameters over the
parameter space of the model. In our \ac{MOND} example, the evidence
for the $\mathcal{S}$ model reads
\begin{equation}
  \begin{split}
    &p(\{\tilde{x}\}\vert \mathcal{S}, I)  = \int \!\!\! \int_{\mathcal{P}}p(\{\tilde{x}\}, k, a_0 \vert \mathcal{S}, I)dk\,da_0  \\
    &= \int \!\!\! \int_{\mathcal{P}}p(\{\tilde{x}\} \vert k, a_0,
    \mathcal{S}, I) p(k,a_{0} \vert \mathcal{S}, I)\,\de k\,\de a_0\,.
    \label{evidence_signal}
  \end{split}
\end{equation}
This is a weighted integral of the likelihood, $p(\{\tilde{x}\}\vert
\partheo_0, \mathcal{S}, I)$ [see Eq.\,(\ref{evidence_signal})], over
the space of unknown parameters, where the weights are set by the
prior distributions of the theory parameters, $k$ and $a_0$ in this
case. The Bayesian evidence thus depends on the volume of the
parameter space and on the priors. If the dimensionality of the
parameter space is large, or if the likelihood and/or the prior are
strongly localized, calculating this integral on a uniform grid in the
parameter space can become computationally costly. A more practical
solution to the problem is to randomly sample the parameter space. To
compute the integral in Eq.\,(\ref{evidence_signal}), we use the
Nested Sampling algorithm, which was specifically designed to
calculate evidence values~\cite{Skilling2004}.

For the $\mathcal{N}$ model, there are no theory parameters to
marginalize over, i.e. the theory parameter space is dimensionless
($\partheo_0=\{\emptyset\}$).  The evidence is thus simply the noise
likelihood,
\begin{equation}
  p(\{\tilde{x}\}\vert \mathcal{N},I) = \prod_{j=1}^{N/2} \frac{1}{\sigma^2_j 2\pi}\exp \left[ -\frac{\vert\tilde{x_j}\vert^2}{2\sigma_j^2}  \right]\,.
  \label{evidence_noise}
\end{equation}

The difference between the likelihoods for models $\mathcal{S}$ and
$\mathcal{N}$, Eqs.\,(\ref{evidence_signal}) and
(\ref{evidence_noise}), respectively, is that in the latter the
Gaussian noise is expressed as $\tilde{n}_j = \tilde{x}_j$, while in
the former $\tilde{n}_j = \tilde{x}_j - \tilde{h}_j$. The likelihood
for model $\mathcal{N}$ can thus be viewed as the likelihood for model
$\mathcal{S}$ with the signal amplitude set to zero. For the Bayes
factor in Eq.\,(\ref{model_selection_ratio_bayes}), the likelihood
normalization terms in cancel out, which simplifies the calculations,
leaving only the exponentials of the likelihoods and the normalization
due to the model priors. The ratio of the model priors represents our
confidence in one model against the other, based on the background
information $I$. In the absence of preference for either model, this
ratio is set to unity, while if background information is available,
it can be included in the prior odds ratio accordingly. We will not
prioritize a model over the other, so that the posterior odds ratio is
simply equal to the Bayes factor.

The posterior odds ratio discussed in this section can be used to
decide whether there was a signal buried in the data gathered during
the \ac{SP} fly-by and to provide a quantitative measure of our
confidence in a signal detection.

\section{Results}
\label{sec_results}

We test our data analysis method on artificially simulated data to
assess the performance of the framework and inspect the various
possible outcomes of the experiment. In order to justify the
experiment feasibility, it is important to establish what conclusions
can be made on the basis of data acquired during the \ac{LPF}
flight. More specifically, we check the implementation of the
parameter estimation and model selection, and determine how well the
parameters values may be inferred and what choices about the model
that best describes the data may be made.

The artificial data is generated following the model defined in
Eq.\,(\ref{signal_model_freq}) and consists of the signal with
additive Gaussian noise characterised by the known \ac{ASD} of the
instrument noise (see Fig.~\ref{LPFnoise}). The real and imaginary
parts of the noise $\tilde{n}(f)$ are treated as statistically
independent and drawn from a Gaussian distribution with the given $\sigma^2 (f)$ providing
\begin{equation}
  \begin{split} 
    p(\tilde{n}(f)) &= p\left(\Re\left[\tilde{n}(f)\right]\right)\,p\left(\Im\left[\tilde{n}(f)\right]\right) \\
    =& \frac{1}{2\pi\sigma^2(f)}\exp\left(-
      \frac{\Re\left[\tilde{n}(f)\right]^{2} +
        \Im\left[\tilde{n}(f)\right]^{2}}{2\sigma^2(f)}\right)\,.
  \end{split}
\end{equation}

For the signal model $\tilde{h}(\parmis_0,\partheo_0)$ we chose a
particular theoretical prediction for the deviations of the gravity
stress tensor from the Newtonian case, as discussed in
Sec.~\ref{sec_TeVeS}.  We test our data analysis setup on \ac{teves},
but we wish to emphasize that this analysis framework is general and
can be used for any signal predictions.

\begin{figure}
  \resizebox{\columnwidth}{!}{%
    \includegraphics[width=\columnwidth]{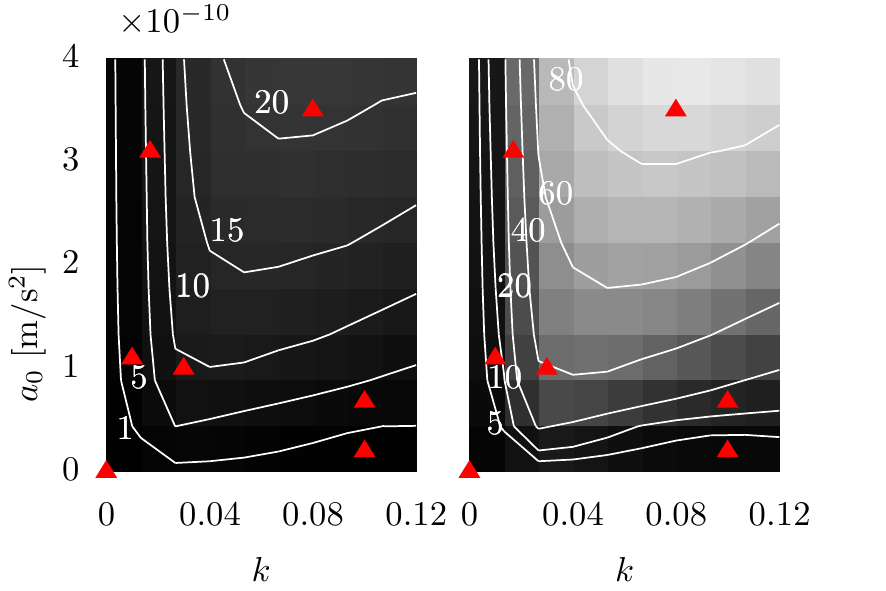}}
  \caption{\ac{SNR} estimates for the current best noise estimate
    (right panel) and the requirements noise (left panel). The
    \ac{SNR}s are calculated at the points in parameter space where
    the \ac{teves} numerical calculations were performed. The 
    triangles correspond to the values of $k$ and $a_0$ for which the
    signal templates were injected into the data (see
    Table~\ref{injection_points}).}
  \label{snr_best_req}
\end{figure}

As shown in Sec.~\ref{subsec_mission_param_analysis}, the mission
parameters can be fixed and do not cause the signal to vary
significantly once they are defined and measured. Throughout the
analysis of the theory parameters, we fix a specific set of mission
parameters values in accordance with Table
\ref{table_lpf_mission_parameters}. We may thus write
\begin{equation}
  \tilde{h}(f_j,\parmis_0,
  \partheo_0) =
  \tilde{h}(f_j,\parmis_0 ,k,a_0) = \tilde{h}(f_j,k,a_0)\,.
\end{equation}
The theory parameter space ($k$, $a_0$) was discussed in
Sec.~\ref{subsection_sm_and_ps} and \ac{SNR}s are calculated following
Eq.\,(\ref{SNR_def}). Figure \ref{snr_best_req} shows the \acp{SNR}
for the chosen \ac{LPF} trajectory as a function of ($k$, $a_0$).  For
large values of both $k$ and $a_0$ the \ac{SNR} reaches values of
$\sim 100$ for the current best estimate and $\sim 20$ for the
requirements noise. This implies that the posterior distributions for
the parameter estimates will be reasonably narrow in those high
\ac{SNR} regions. Conversely, we expect signals residing in low
\ac{SNR} areas to have correspondingly broader posterior
probabilities.

\begin{table}
  \caption{Values of $k$ and $a_0$ for which the signal template was
    injected in the data to probe parameter estimation.}
  \label{injection_points}
  \begin{tabular}{ccc}
    \toprule[1.pt]
    \toprule[1.pt]
    \addlinespace[0.3em]
    Number & $k$ & $a_0$ [$10^{-10}\,\text{m/s}^2$] \\
    \addlinespace[0.2em]
    \midrule[1.pt]
    \addlinespace[0.2em]
    $1$ & $ 0.030$ & $1.00$ \\
    $2$ & $ 0.080$ & $3.50$ \\
    $3$ & $ 0.010$ & $1.10$ \\
    $4$ & $ 0.017$ & $3.10$ \\
    $5$ & $ 0.100$ & $0.20$ \\  
    $6$ & $ 0.100$ & $0.68$ \\ 
    $7$ & $0$ & $0$ \\
    \bottomrule[1.pt]
    \bottomrule[1.pt]
  \end{tabular}
\end{table}

Given the \ac{SNR} estimates shown in Fig.~\ref{snr_best_req}, we
choose a number of representative points in the parameter space with
high, intermediate, and low \ac{SNR} values, and estimate their
posterior probabilities. These points are listed in
Table~\ref{injection_points}. We start with \textit{point 1}, for
which $k$ and $a_0$ take their ``standard'' values~\cite{Neil}. This
point belongs to the high \ac{SNR} region. To test the area with the
loudest \acp{SNR}, we probe \textit{point 2}. A third interesting
region, where the performance of our interpolation must be checked, is
the area near the boundary that was imposed on the prior parameter
space [Eq.\,(\ref{equation_boundary})]. We chose two points here:
\textit{point 3} and \textit{point 4} for low and high \acp{SNR},
respectively. Further, we consider two points with low \acp{SNR}:
\textit{point 5} and \textit{point 6}. They are chosen relatively
close to each other in order to assess the area where the transition
from the detectable to nondetectable signal might occur. Finally, we
consider \textit{point 7}, where the Newtonian limit of the theory
lies and we expect to find no signal in the data. For each chosen
point on the parameter space we perform $200$ simulations with
different noise realizations.

\subsection{Parameter estimation}
The experiment can give us insight into how well the parameters of the
theory can be recovered and constrained from the data. This can be
achieved by calculating the posterior probability distribution for the
parameters.  We have an initial prior assumption for the parameter
values, which in our case is a simple uniform distribution over the
predefined parameter space discussed in
Sec.~\ref{subsection_sm_and_ps}. We compute evidence values using a
random sampling algorithm (Nested Sampling ~\cite{Sivia,Skilling2006})
as a mean to overcome potential issues due to the sampling of the
theory parameter space, or to its high dimensionality. While the
theory parameter space is two-dimensional in our example, we must be
ready to consider theories with a higher number of parameters. The
algorithm and its specific implementation we used,
\verb+MultiNest+~\cite{Multinest}, are designed to efficiently sample
a parameter space and to output the samples from the joint posterior
parameter distribution and the Bayesian evidence.

\begin{figure*}
  \resizebox{0.72\textwidth}{!}{%
    \includegraphics[width=\columnwidth]{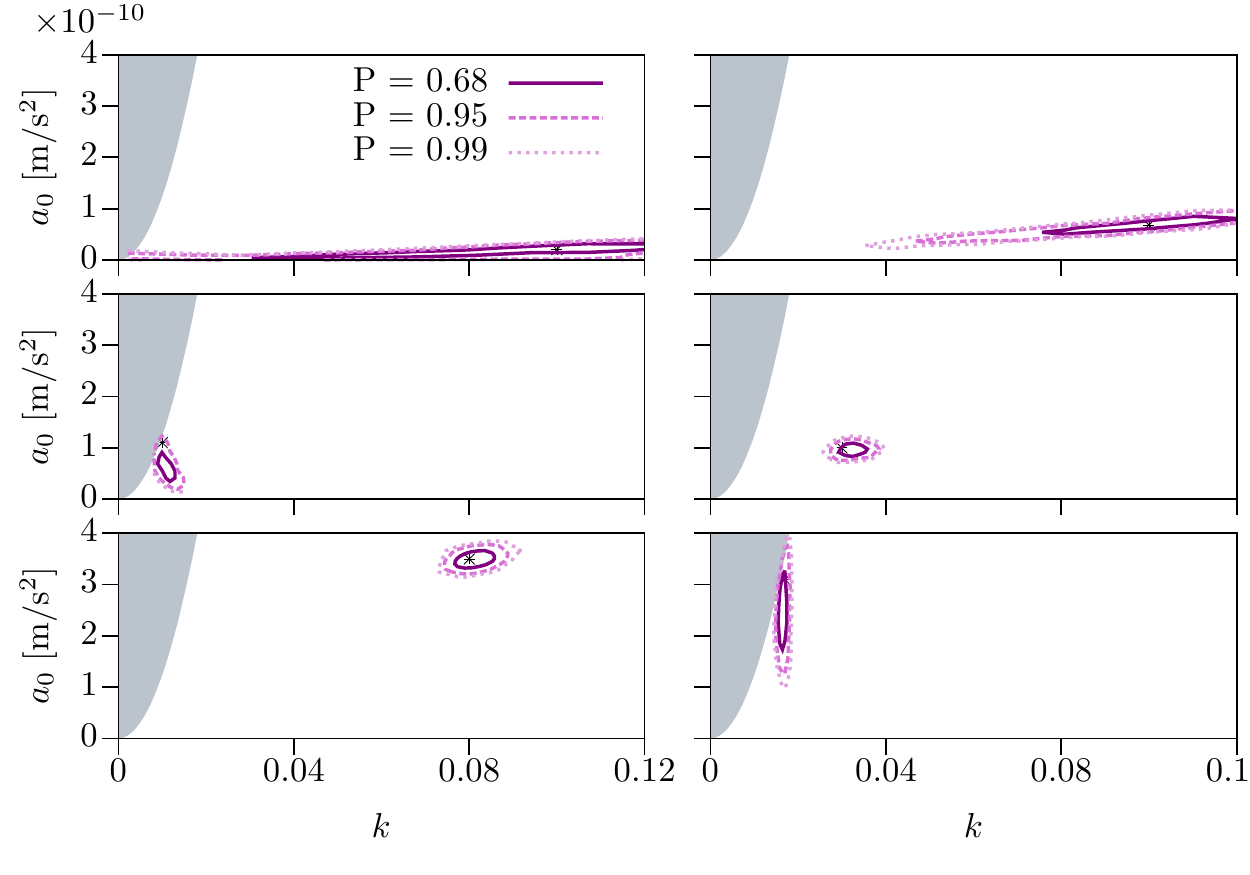}}
  \caption{Joint posterior probability distribution for the parameters
    $k$ and $a_0$ using the current best estimate noise
    model. Contours represent lines of constant probability density
    defining regions that enclose $68$\%, $95$\%, and $99$\% of the
    probability. The panels represent $6$ signal injections at the
    first $6$ points in the parameter space listed in Table
    \ref{table_sigma_k_a0}.}
  \label{contours}
\end{figure*}

\begin{figure*}
  \resizebox{0.72\textwidth}{!}{%
    \includegraphics[width=\columnwidth]{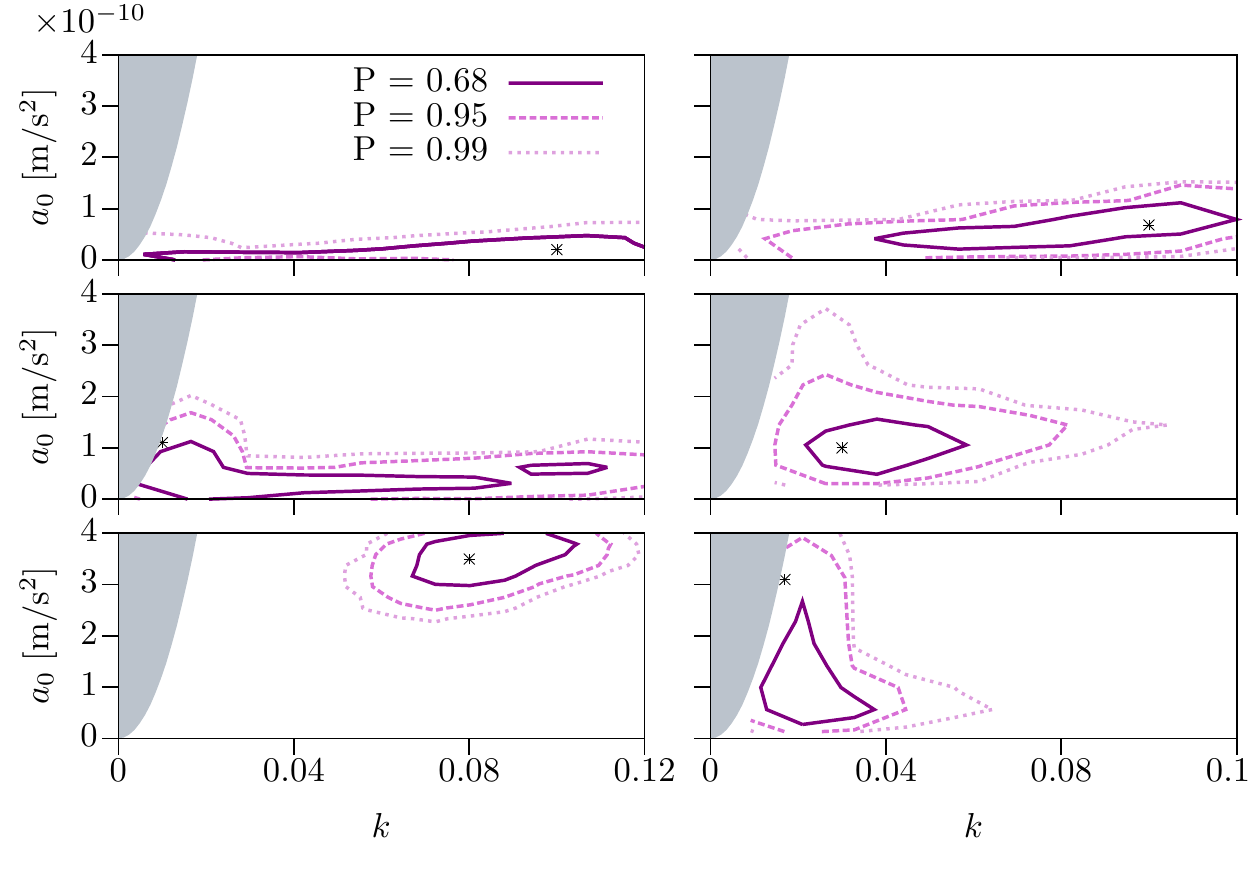}}
  \caption{Same as Fig.\,\ref{contours} but for the requirements noise
    model.}
  \label{contours_req}
\end{figure*}

To quantitatively summarize the information on the posterior
distributions of the parameters, it is natural to use confidence
intervals. These indicate the parameter range within which the area
enclosed under the posterior has a certain probability. This provides
an estimate on how confident we are that the value of a parameter
falls in that range. As is customary, use the confidence interval
values $68$\%, $95$\%, and $99$\%, which correspond to $1\sigma$,
$2\sigma$, and $3\sigma$ deviations of a parameter from its mean value
in the special case of a one dimensional Gaussian
distribution. Accordingly, we define the confidence contours
\begin{equation}
  \begin{split}
    P((k,a_0) \subset S,I ) & = \int \!\!\! \int_{S} p(k,a_0 \vert \{x \}, I )\,\de k\,\de a_0  \\
    & = (68\%, 95 \%, 99 \%),
  \end{split}
\end{equation}
where the space $S$ corresponds to the minimal volume underneath the
posterior probability that integrates to predefined probability. The
resulting contours also represent lines of constant probability
density. Figure \ref{contours} shows the contour plots of the joint
posterior distributions for the parameters $k$ and $a_0$ for simulated
signals located at selected parameter space positions.

\begin{table}[!t]
  \caption{Average values of the standard deviations $\overline{\Delta k}$ and $\overline{\Delta a_0}$ of the one dimensional posteriors of the parameters. The values are given for the $6$ points in the $(k, a_0)$ parameter space where the {\it true} signal injections were made. The averages are determined from $200$ different noise realisations (using the current best estimate noise) and posterior estimates truncated by our priors are artificially reduced.}
  \resizebox{\columnwidth}{!}{%
    \begin{tabular}{cc cc cc}
      \toprule[1.pt]
      \toprule[1.pt]
      \addlinespace[0.3em]
      & & \multicolumn{2}{c}{Current best estimate} & \multicolumn{2}{c}{Requirements noise}  \\  \midrule[1.pt]
      \addlinespace[0.2em]
      $k$ & $a_0$ & $\overline{\Delta k}$ & $\overline{\Delta a_{0}}$ & $\overline{\Delta k}$ & $\overline{\Delta a_{0}}$ \\
      & [$10^{-10}\,\text{m/s}^2$]  & &  [$10^{-10}\,\text{m/s}^2$] & & [$10^{-10}\,\text{m/s}^2$] \\
      \addlinespace[0.2em]
      \midrule[1.pt]
      \addlinespace[0.2em]
      0.030 & 1.00 & 0.00203 & 0.096 & 0.0121 & 0.687 \\
      0.080 & 3.50 & 0.00306 & 0.117 & 0.0125 & 0.352 \\
      0.010 & 1.10 & 0.00087 & 0.225 & 0.0295 & 0.515 \\
      0.017 & 3.10 & 0.00066 & 0.422 & 0.0066 & 0.907 \\
      0.100 & 0.20 & 0.03053 & 0.084 & 0.0345 & 0.173 \\
      0.100 & 0.68 & 0.01838 & 0.137 & 0.0295 & 0.268 \\
      \bottomrule[1.pt]
      \bottomrule[1.pt]
    \end{tabular}
  }
  \label{table_sigma_k_a0}
\end{table}

The resulting estimates of the posterior probabilities are shown in
Figs.~\ref{contours} and~\ref{contours_req} for the current best
estimate noise and for the requirements noise, respectively. The
results are presented for a single noise realization. Estimates for
the standard deviation of the posterior distributions of $k$ and $a_0$
averaged over $200$ noise realizations for the current best noise
estimate and requirements noise are given in
Table~\ref{table_sigma_k_a0}. For signals with high \acp{SNR} (see
Fig.~\ref{snr_best_req}) the posterior likelihoods are narrow and
exhibit low correlation between the two parameters. This means that in
the case of signal detection it would be possible to estimate them
with relatively small uncertainties. For lower \acp{SNR}, however, the
error on $k$ is much larger than one on $a_0$. In some cases the error
on $k$ is limited only by the range of the parameter prior. This
scenario will be considered in more detail in
Sec.\,\ref{no_signal_injection}, which is dedicated to the case of
noise-only simulated data.
 
\begin{figure}[!b]
  \resizebox{\columnwidth}{!}{%
    \includegraphics[width=\columnwidth]{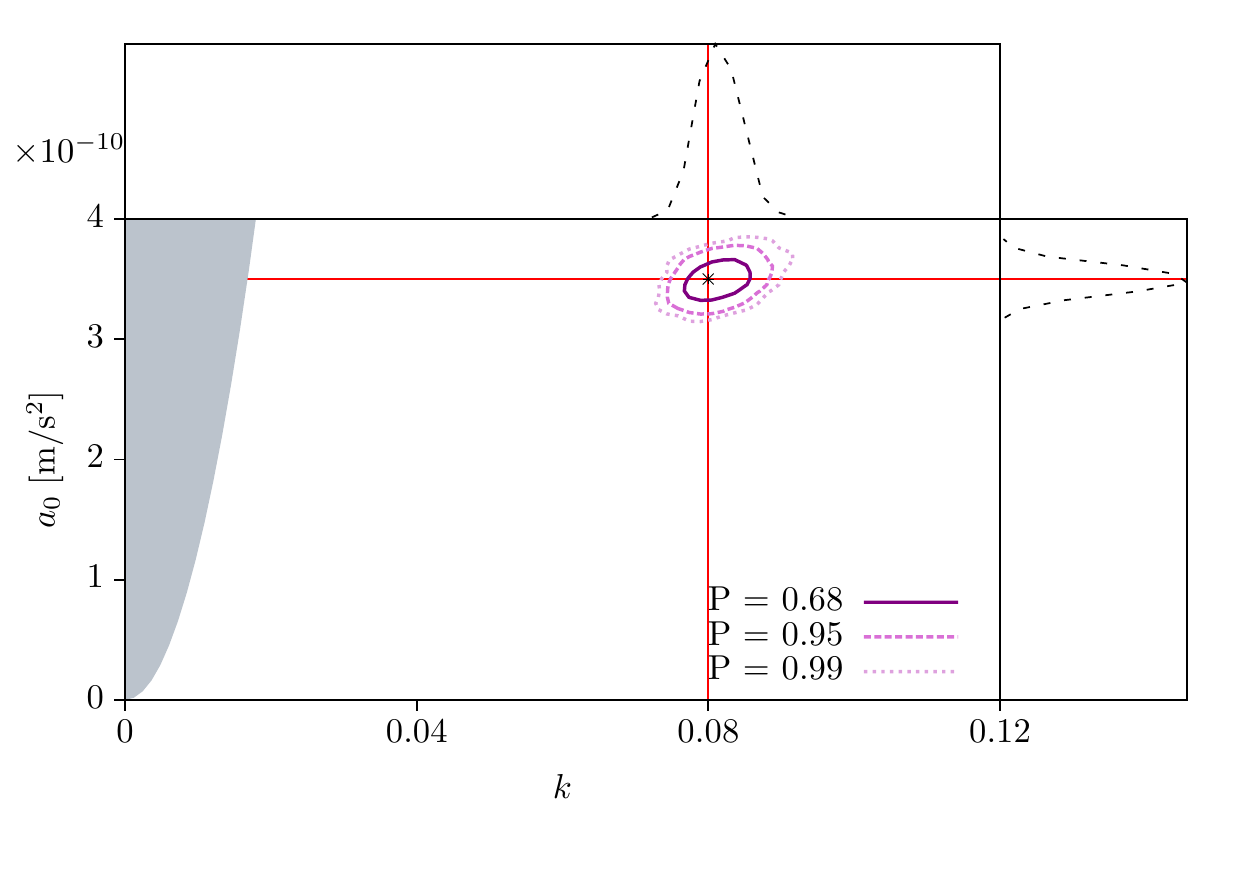}}
  \caption{Posterior probability distributions and marginalised
    posterior distributions for the current best noise estimate for
    the parameters of the injected signal at $k = 0.08$ and $a_0 = 3.5
    \cdot 10^{-10} \accunits$. The red lines indicate the true values
    at which the simulated signal was injected.}
  \label{marg_5}
\end{figure}
\begin{figure}[!b]
  \resizebox{\columnwidth}{!}{%
    \includegraphics[width=\columnwidth]{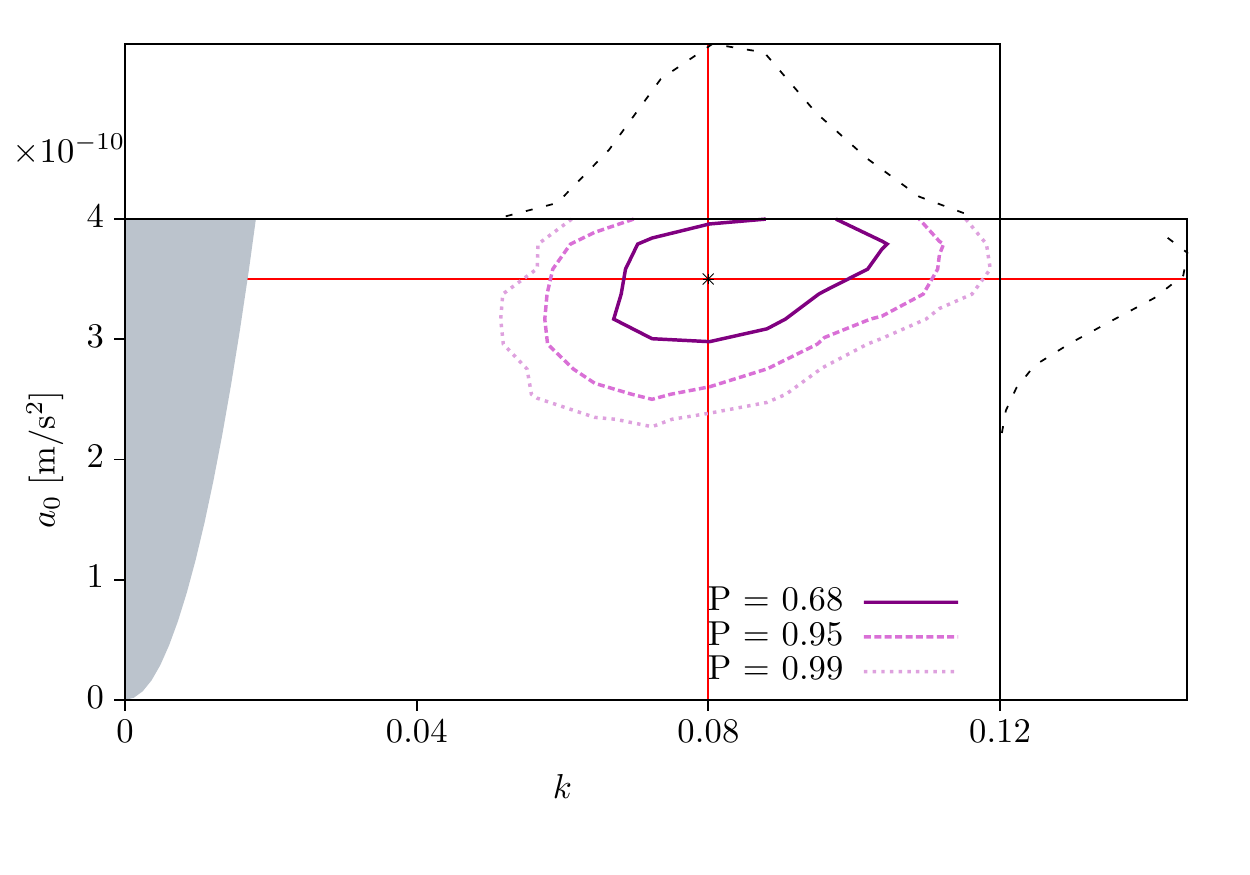}}
  \caption{Posterior probability and marginalised posterior
    distributions for requirements noise for the parameters of the
    injected signal at $k = 0.08$ and $a_0 = 3.5 \cdot 10^{-10}
    \accunits$. The red lines indicate the true values at which the
    simulated signal was injected.}
  \label{marg_5_req}
\end{figure}

\begin{figure}[!t]
  \resizebox{\columnwidth}{!}{%
    \includegraphics[width=\columnwidth]{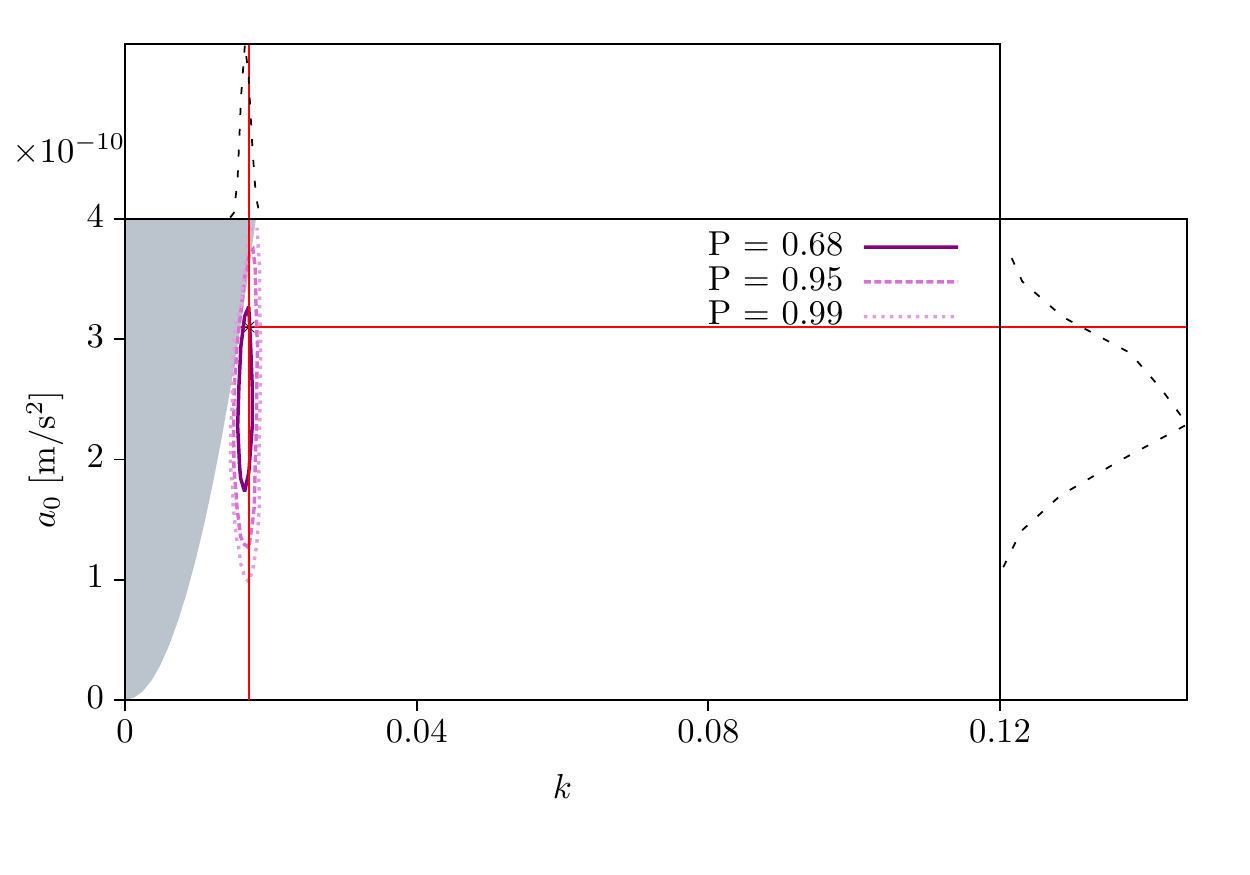}}
  \caption{Posterior probability distributions and marginalised
    posterior distributions for the current best noise estimate for
    the parameters of the injected signal $k = 0.017$ and $a_0 = 3.1
    \cdot 10^{-10} \accunits$. The red lines indicate the true values
    at which the simulated signal was injected.}
  \label{marg_6}
\end{figure}

\begin{figure}[!t]
  \resizebox{\columnwidth}{!}{%
    \includegraphics[width=\columnwidth]{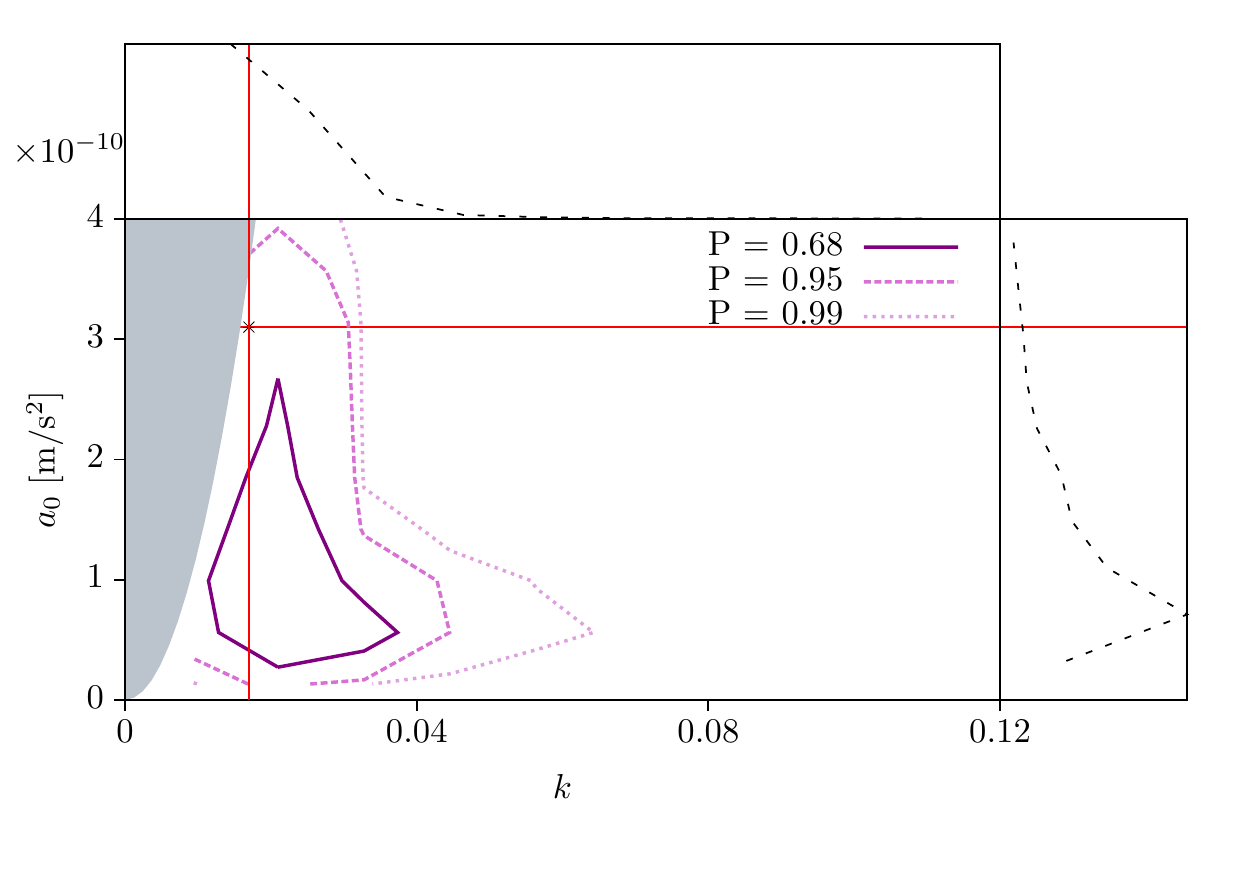}}
  \caption{Same as Fig.~\ref{marg_6} but for the requirements noise.}
  \label{marg_6_req}
\end{figure}

\begin{figure}[!t]
  \resizebox{\columnwidth}{!}{%
    \includegraphics[width=\columnwidth]{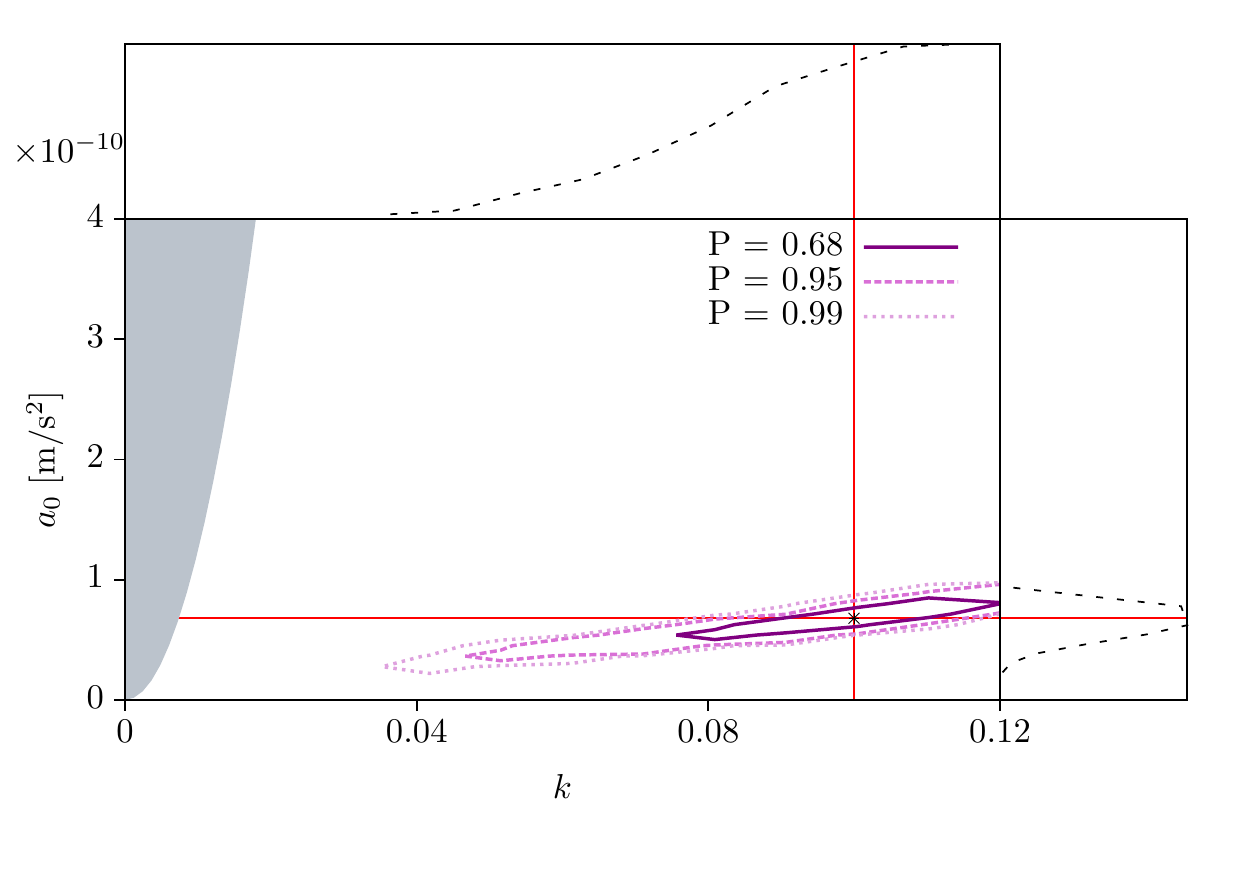}}
  \caption{Posterior probability distributions and marginalised
    posterior distributions for the current best noise estimate for
    parameters of the signal modelled for $ k = 0.1$ and $a_0 = 0.68
    \cdot 10^{-10} \accunits$. The red lines indicate the true values
    at which the simulated signal was injected.}
  \label{marg_2}
\end{figure}

\begin{figure}[!t]
  \resizebox{\columnwidth}{!}{%
    \includegraphics[width=\columnwidth]{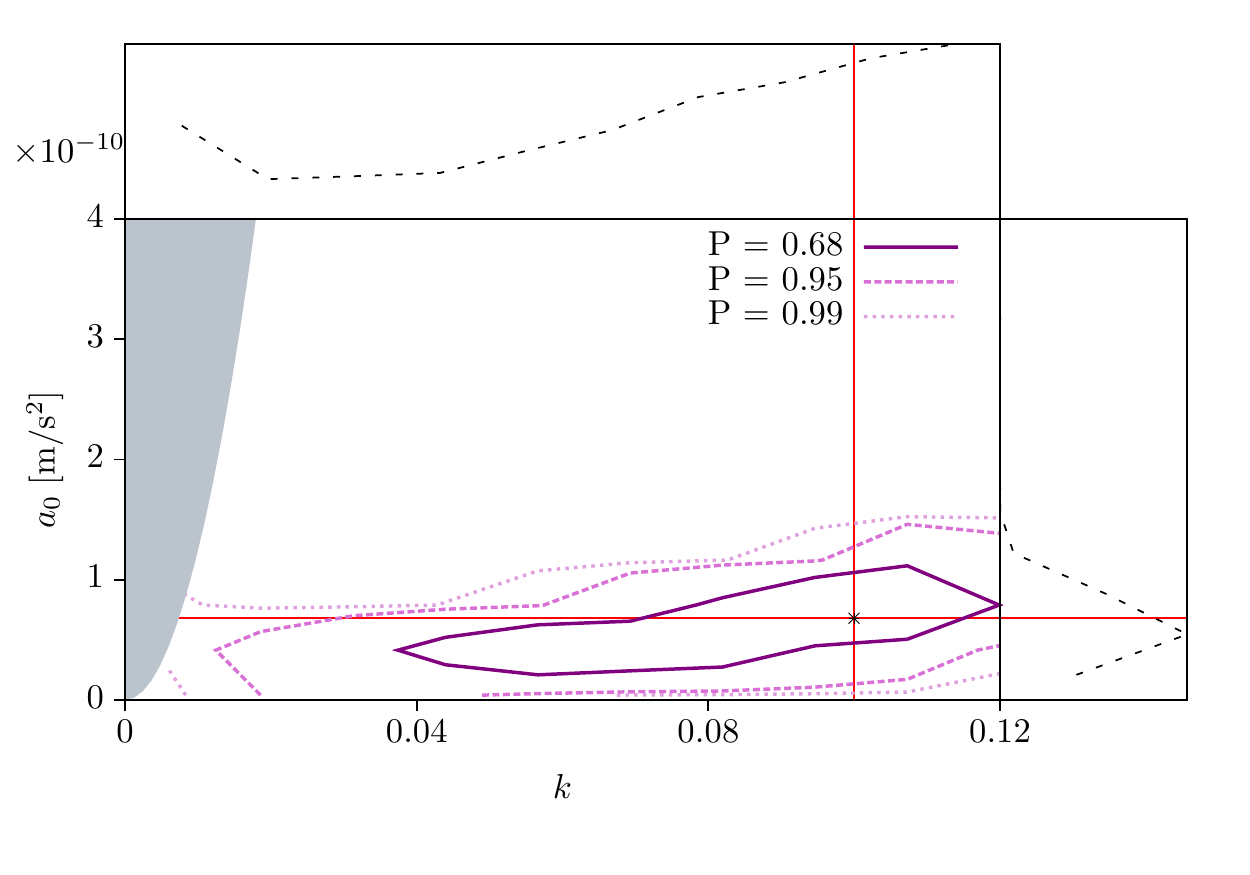}}
  \caption{Same as Fig.~\ref{marg_2} but for the requirements noise.}
  \label{marg_2_req}
\end{figure}

Using Eqs.\,(\ref{k_distribution}) and (\ref{a0_distribution}) we
determine the marginal distributions for the parameters $k$ and $a_0$
and their expected values. These marginalized posterior distributions
allow us to identify three types of results within our six signal
simulations. As shown in Figs.\,\ref{marg_5} and \ref{marg_5_req}, for
the first type of result the joint posterior distribution is narrow
and well localised, especially for the current best estimate noise. In
this scenario the marginal distributions of both $k$ and $a_0$ can be
estimated relatively well. Results for the second case can be found in
Figs.\,\ref{marg_6} and~\ref{marg_6_req}.  This time the posterior is
near the boundary of the prior established in
Sec.~\ref{subsection_sm_and_ps}. The uncertainty on $a_0$ is much
broader than the one on $k$. Finally, Figs.\,\ref{marg_2} and
\ref{marg_2_req} show the third kind of result: the marginalized
distribution for $k$ is very broad and is determined by the range that
was imposed on it as a prior. In this low \ac{SNR} regime, it will be
hard to make estimates for $k$.



\subsection{The no signal injection case}
\label{no_signal_injection}

No deviations from Newtonian gravity potential have been observed so
far in the Solar System. Hence, this is a particularly important case
for our analysis and corresponds to a dataset containing no signal. We
consider this case as a likely outcome of the experiment and wish to
assess the impact that a measurement of data with no signal would have
on the theory parameter space, i.e. which observation-based
restrictions can be placed on the $(k, a_0)$ space.

In Figs.\,\ref{marg_2} and \ref{marg_2_req}, we already saw the shape
of the posterior distribution in the case of low \acp{SNR}. We would
expect to have somewhat similar results for the case of a noise-only
data model, i.e.~when we set $\tilde{h}(f_j,\parmis_0,\partheo_0) = 0$
in Eq.\,(\ref{signal_model_freq}). On the basis
 of the theory proposed
in~\cite{TeVeS}, we place the Newtonian limit of the theory at $a_0 =
0$, thus setting the gravity stress tensor to be equal to the
Newtonian stress tensor for all templates on the $k$-axis.

\begin{figure*}
  \resizebox{0.72\textwidth}{!}{%
    \includegraphics[width=\columnwidth]{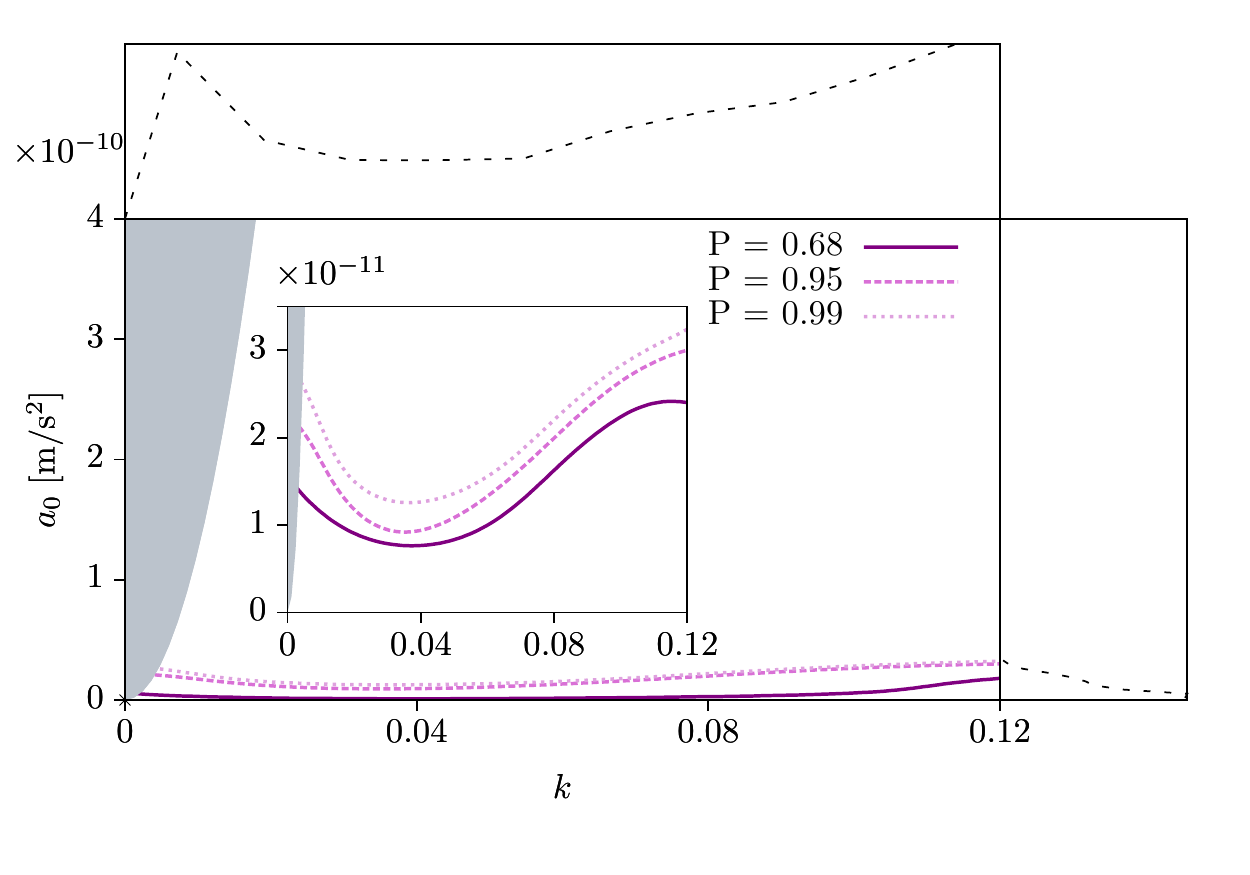}}
  \caption{Posterior probability density for the current best estimate
    noise realisation in the case of no signal injection,
    i.e.~$\tilde{h}(f_j,\parmis_0,\partheo_0) = 0$.}
  \label{no_sig_inj}
\end{figure*}

\begin{figure*}
  \resizebox{0.72\textwidth}{!}{%
    \includegraphics[width=\columnwidth]{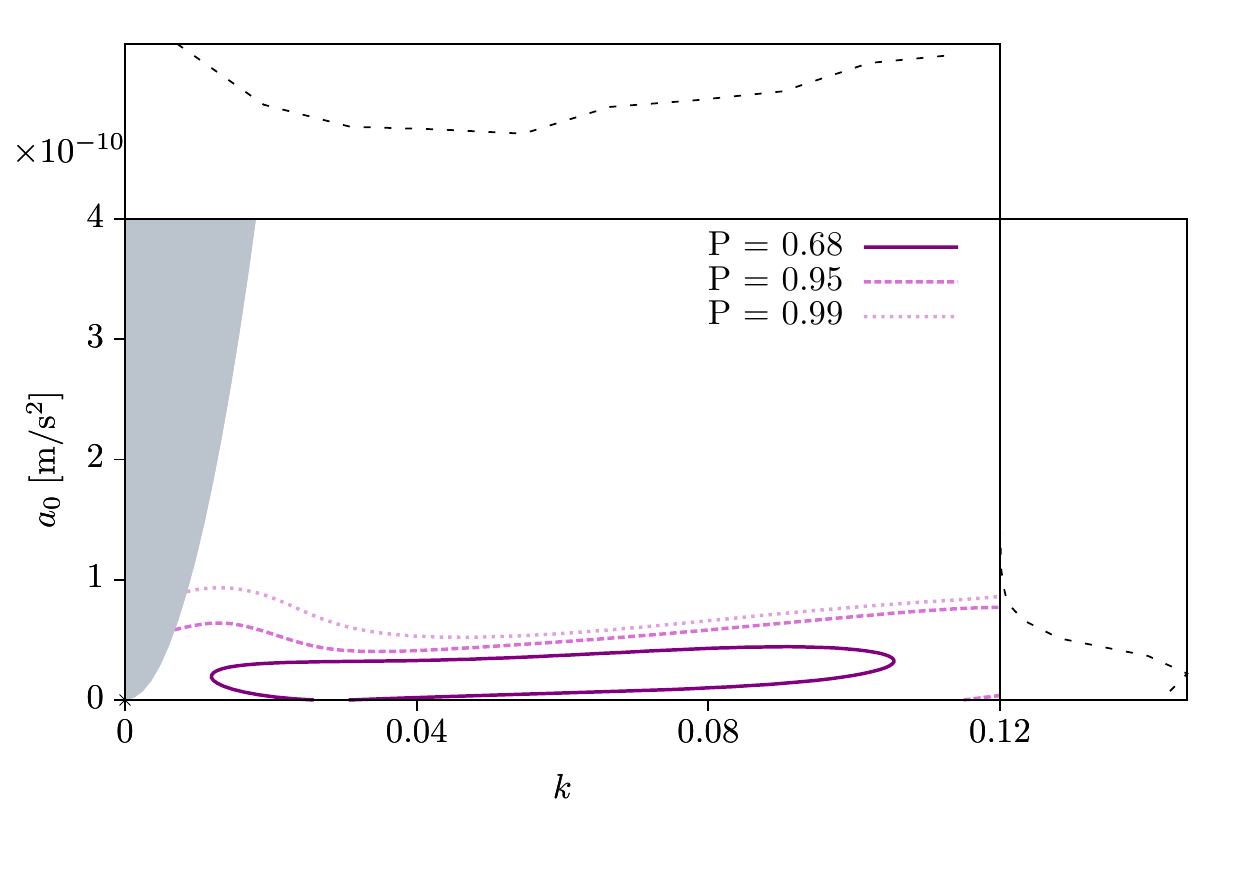}}
  \caption{Same as Fig.~\ref{no_sig_inj} but for the requirements
    noise.}
  \label{no_sig_inj_req}
\end{figure*}

We perform $200$ simulations, each with a different noise realization,
for both the current best estimate and requirements noise models.  We
determine $68\%$, $95\%$, and $99\%$ confidence interval for both of
them. To visualise the restriction on the parameter space that
follows, we chose a representative noise realization. The results in
Figs.\,\ref{no_sig_inj} and \ref{no_sig_inj_req} show uncertainty on
the determination of the parameter $k$, meaning that a null
measurement would not help us constrain $k$ at all, whereas $a_0$
would be tightly bounded.  We note, however, that the point in the
  parameter space for the standard choice of parameters $k = 0.03$ and
  $a_0 = 10^{-10}\, \accunits$ would be ruled out.  The average error
on the marginalized posterior distribution of $a_0$ for the current
best noise estimate is $\overline{\Delta a_0} = 0.055\cdot 10^{-10}\,
\accunits$, while for the requirements noise it is $\overline{\Delta
  a_0} = 0.154\cdot 10^{-10}\, \accunits$.

\subsection{Model Selection}
\label{results_model_selection}

\begin{figure*}[!t]%
  \resizebox{0.72\textwidth}{!}{%
    \includegraphics[width=\columnwidth]{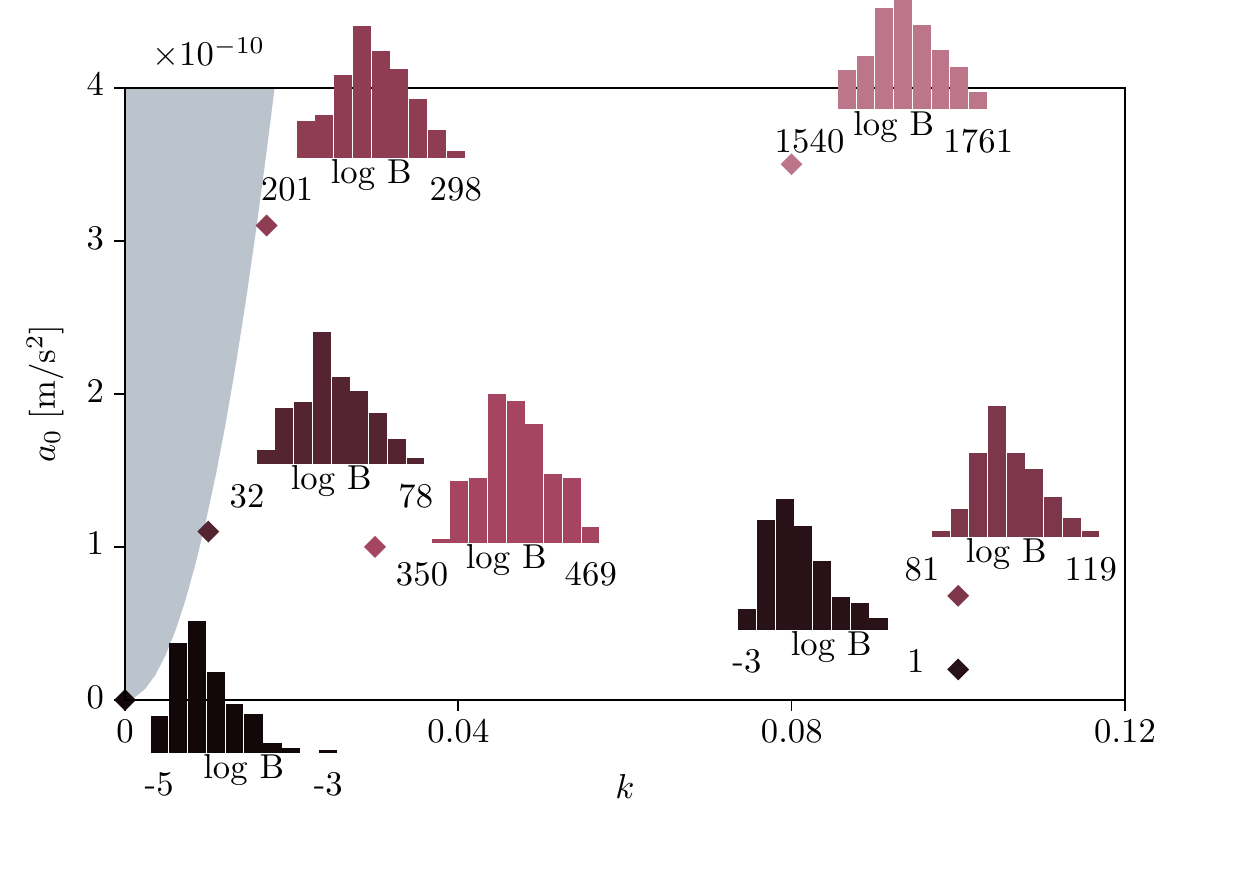}}%
  \caption{Histograms of the logarithms of the Bayes factor $\log B$
    for the $200$ noise realisations (current best estimate) at the
    $7$ representative points in the parameter space, where the
    signals were injected. These points are listed in Table
    \ref{injection_points}.}
  \label{bayes_factors_best}
\end{figure*}

We now follow Eq.\,(\ref{model_selection_ratio_bayes}) and compute the
Bayes factor\footnote{We remind the reader that we set the prior model
  odds to unity.} between our two candidate models $\mathcal{S}$ and
$\mathcal{N}$ using the signals calculated for the sets of parameters
listed in Table \ref{injection_points}. This gives a measure of the
signal detectability in noise, depending on the combination of the
theory parameters $\partheo_0=\{k,a_0\}$, allowing us to quantify the
confidence in one model relative to the other on the basis of the
outcome of the experiment. As discussed in Sec.~\ref{model_selection},
the $\mathcal{S}$ hypothesis assumes that the data is the sum of noise
and a signal that depends on $k$ and $a_0$, while the $\mathcal{N}$
hypothesis assumes it to be noise-only and to have no parameter
dependencies. As indicated in Eq.\,(\ref{evidence_signal}), the
$\mathcal{S}$ hypothesis requires us to integrate the joint
probability $p(\{\tilde{x}\},k,a_0)$ over the parameter space of the
signal $(k,a_0)$, whereas the evidence for the noise-only model is
simply given by the likelihood in Eq.\,(\ref{evidence_noise}).

\begin{figure*}
  \resizebox{0.72\textwidth}{!}{%
    \includegraphics[width=\columnwidth]{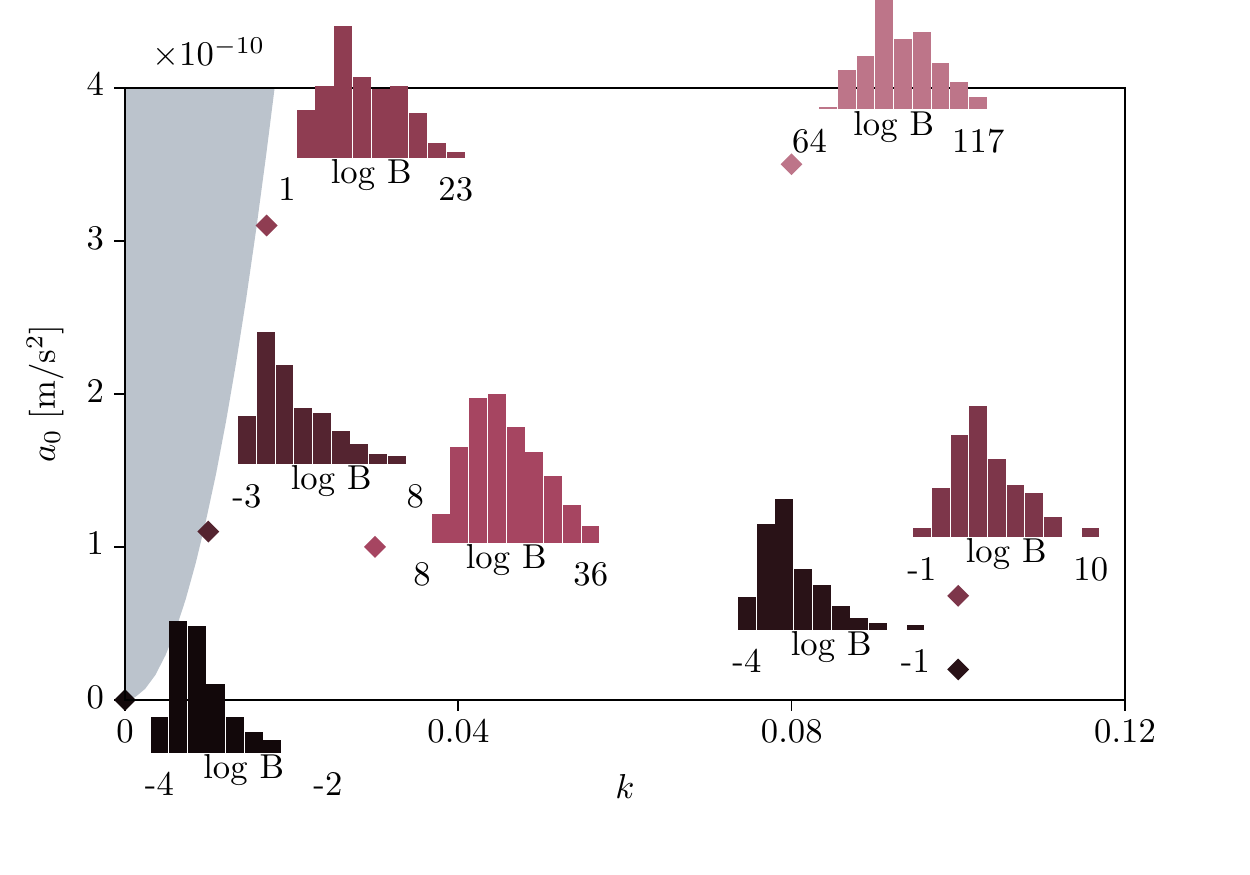}}%
  \caption{Same as Fig.~\ref{bayes_factors_best} but for the
    requirements noise.}
  \label{bayes_factors_req}%
\end{figure*}

In reality, we will have a single measurement yielding one value for
the Bayes factor which itself is a random variable subject to
variations between noise realisations. By performing an analysis of
the artificial data, however, we can study the distribution of the
Bayes factor and therefore understand the interpretation of a single
value measurement. For the model selection we analysed the same data
as for the parameter estimation. The Bayes factors distributions
dependence upon the theory parameters is found in
Fig.~\ref{bayes_factors_best} for the current best estimate noise
model and in Fig.~\ref{bayes_factors_req} for the requirements
noise. We show the logarithms of the Bayes factor estimates at the $7$
representative points in the parameter space collected in Table
\ref{injection_points}. In $5$ cases out of $7$ the Bayes factor
logarithms all have positive values: this means that the $\mathcal{S}$
hypothesis will be strongly favoured over the $\mathcal{N}$
hypothesis. On the other hand, negative logarithms of the Bayes factor
imply that the noise-only model $\mathcal{N}$ is favoured. This occurs
in $2$ cases out of $7$. One of these is the noise-only $({k = 0}, \;
{a_0 = 0 \; \accunits}$) point, where the data only contains noise:
this behaviour is therefore expected. The second point is at $({k =
  0.1}, \; {a_0 = 0.2 \cdot 10^{-10} \; \accunits})$.  In this case,
noise and signal are mixed, but a rejection of the $\mathcal{S}$
hypothesis is likely.

The analysis just discussed shows a rigorous way of determining the
detectability of a signal. While we solely considered a noise-only
model and a signal model of \ac{MOND}ian inspiration, we note that our
analysis can be extended to include other models, as, for example,
models with non-Gaussian noise or ones incorporating glitches that
could resemble the signal. In addition we can probe whether the data
will be best described by one theory or another when it exhibits a
deviation from the Newtonian background.

\subsection{Detector noise artifacts}

So far we analysed the simulated \ac{LPF} data with noise taken to be
Gaussian and \ac{ASD} defined by the theoretical amplitude spectral
density of \ac{LPF}. In reality, however, non-Gaussian glitches might
appear in the noise as shown in the measurement of the differential
displacements from the test campaigns for \ac{LPF} \footnote{The
  \ac{LPF} spacecraft is already being prepared for launch and is
  undergoing several instrumental tests. To assess the impact of noise
  artifacts, we took the data available from the \ac{LPF} \ac{OSTT}
  performed by Astrium Ltd., Astrium Satellite GmbH (\ac{ASD-astrium})
  extensively testing the end-to-end performance of the \ac{OMS}.
  However, we would like to point out that the noise artifacts might
  have been artificially caused by the test environment.}
\cite{Hechenblaikner2013,Hechenblaikner2010, OSTT_test_report}.  We
now examine the response of our data analysis framework to glitches by
performing parameter estimation and model selection on the \ac{OSTT}
data. We keep working in the \ac{teves} $(k, a_0)$ parameter space and
use the signal templates produced within this theory.

\begin{figure}[!t]
  \resizebox{0.9\columnwidth}{!}{%
    \includegraphics[width=\columnwidth]{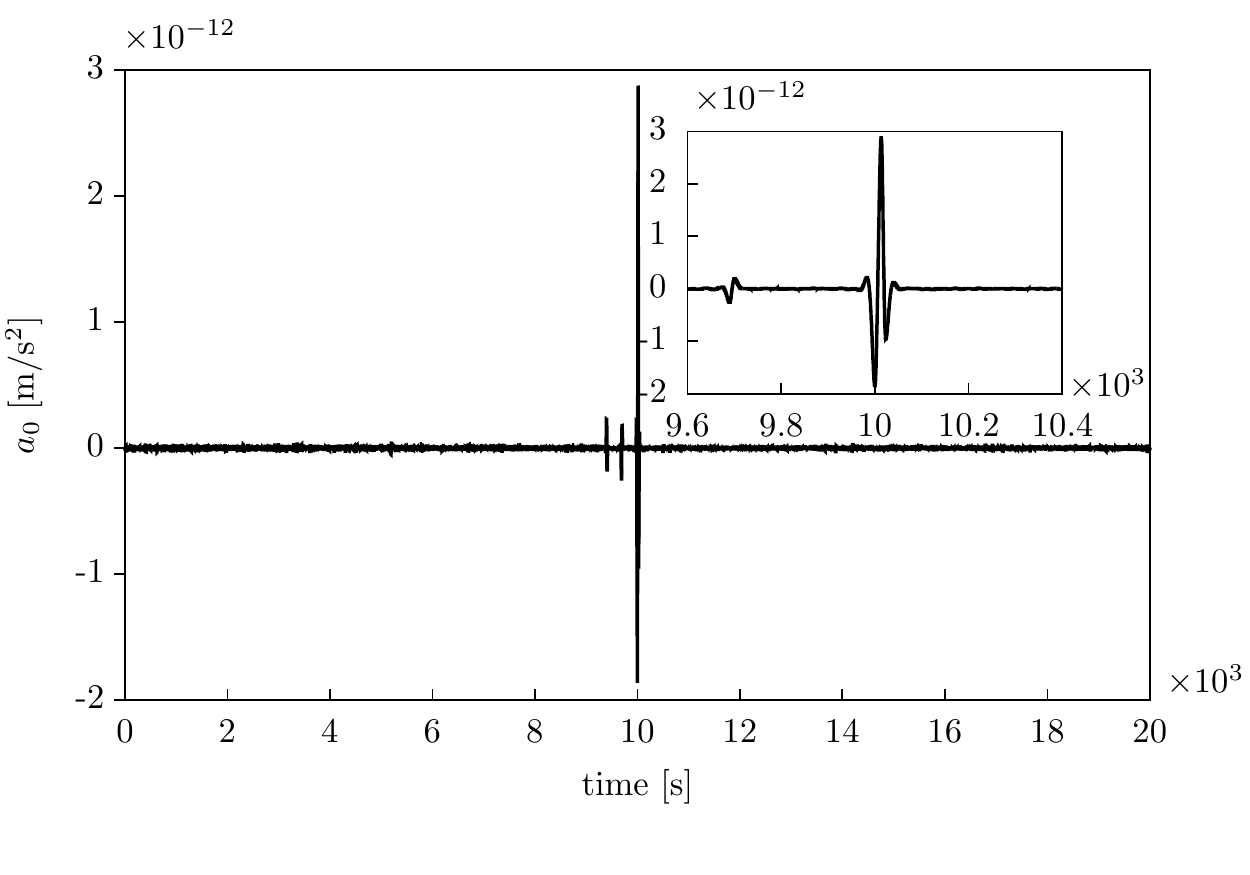}}
  \caption{Data from the test campaign with a glitch of unknown origin
    bandpass filtered in the sensitive frequency band of \ac{LPF}.}
  \label{glitch_twoplots}
\end{figure}

\begin{figure}[!t]%
  \resizebox{\columnwidth}{!}{%
    \includegraphics[width=\columnwidth]{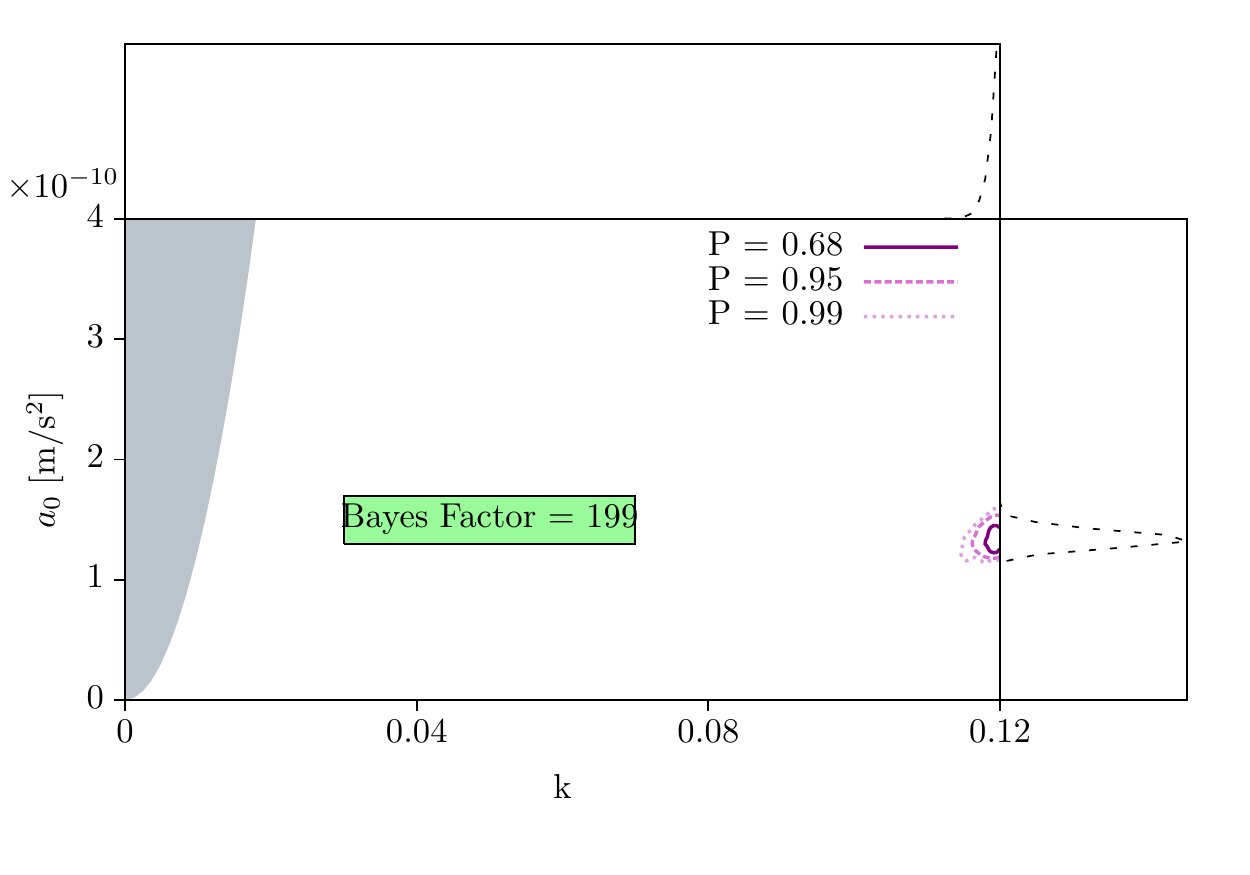}}
  \caption{Resulting posterior probability density for the parameter
    estimation in case of the realistic data of
    Fig.~\ref{glitch_twoplots}.}%
  \label{glitch_0}%
\end{figure}

We shift the test campaign data so that a glitch occurs at the
expected signal arrival time, as shown in
Fig.\,\ref{glitch_twoplots}. We then estimate the posterior
probability distribution for $k$ and $a_0$ for this dataset. Results
are presented in Fig.\,\ref{glitch_0}. The posterior probability peaks
at $(k = 0.12, a_0 = 1.34 \cdot 10^{-10}$ m/s$^2)$.  The standard
deviations for the two parameters are given by $\Delta k = 0.001$ and
$\Delta a_0 = 0.07 \cdot 10^{-10}$ m/s$^2$, respectively. The
recovered parameter values are in the parameter space region that is
inconsistent with the noise-only model. Additionally, the estimated
value of the parameter $k$ is on the boundary of the parameter range
defined by the parameter priors.

The logarithm of the Bayes factor is $\log p(\mathcal{S} \vert
\{\tilde{x}\},I)/p(\mathcal{N} \vert \{\tilde{x}\}, I) = 199$, so that
the $\mathcal{S}$ hypothesis is prioritised over the $\mathcal{N}$
one. This can happen if the characteristic frequency of the glitch is
similar to the characteristic frequency of the signal and highlights
that, in order to achieve confident signal detection, we must
introduce more realistic noise models. In particular, these should
describe non-Gaussianities in the noise, such as glitches. With such
noise models it would be possible to extend the model selection
described in Sec.~\ref{model_selection} and distinguish between noise
artifacts and authentic signals.  The question of the
non-stationarities and glitches in the data is particularly important
in the setup of this experiment because our measurement relies on one
or two repetitions at the most (one or two \ac{SP} fly-by's).
Multiple \ac{SP} fly-by's can significantly increase our confidence in
signal detection against glitches in the data.  However,
distinguishing between noise glitches and signal, and characterising
glitches are very important topics that will need further
investigation.

\section{Conclusions and Future Work}
\label{sec_conclusions}

In this paper we developed a data analysis approach to test
alternative theories of gravity with \ac{LPF}. As shown in
Eq.\,(\ref{stress_projection}), the gravitational stress tensor
affects the relative acceleration between the two test masses onboard
the spacecraft. The tidal field can be sampled by \ac{LPF}, allowing
us to measure its (dis)agreement with the Newtonian tidal field. The
time series that an \ac{LPF} measurement will provide depends on the
trajectory of the spacecraft and on the orientation of its sensitive
axis via the seven mission parameters listed in
Eq.\,(\ref{lpf_parameters}). The data analysis framework we built will
allow for quantitative statements on measuring the tidal field and
posing constraints on alternative theories of gravity.

Testing our data analysis approach required picking a theory of
gravity that predicts deviations from the Newtonian tidal stresses
within the Solar System, where \ac{LPF} will fly. As discussed in
Sec.~\ref{sec_TeVeS}, we considered the example of the \ac{teves}
theory. This choice is convenient as we are able to calculate signal
templates from it. In the regime of our interest, the signal measured
by \ac{LPF} depends on two theory parameters only, namely, a
dimensionless coupling constant $k$ and a characteristic acceleration
$a_0$. Having picked an alternative theory of gravity, we were able
to quantify how the signal is influenced by variations of each of the
mission parameters. We concluded that, within the errors on the
measurement of the position of the spacecraft, the variations of the
signal will be negligible. This is a crucial result as it allows us to
fix the values of the mission parameters when building signal
templates in order to carry out a Bayesian analysis of the theory
parameter estimation and a model selection.  However, the impact of
possible correlations between the mission and the theory parameters on
the conclusions drawn so far in our study was not assessed.
Nevertheless, we expect this correlation to be insignificant and leave
this investigation for the future work.

The results of our Bayesian analysis are presented in
Sec.~\ref{sec_results} in the form of posterior distributions for the
two theory parameters that determine the signal. These are obtained by
considering $200$ different noise realisations. Some combinations of
the parameters (\textit{point} 1 and \textit{point} 2 in Table
\ref{injection_points}) yield a sharp and narrow joint posterior
distribution, indicating that it will be possible to estimate the
theory parameters with high precision in case of high \acp{SNR}.  For
weak signals with low \acp{SNR} (\textit{point} 5 and \textit{point} 6
in Table \ref{injection_points}) the parameter $k$ can only be poorly
estimated from the posterior probability. The results for the current
best estimate of the noise systematically exhibit better parameter
estimation and better distinction between the noise and the signal
hypotheses (see Figs.~\ref{bayes_factors_best}
and~\ref{bayes_factors_req}) than the requirements noise. As the
former model was built upon the estimates of the noise from the flight
hardware test campaigns (see Sec.~\ref{sec:2noises}), it is a good
approximation of the noise during flight.

We also considered the special case in which the data consists of
noise only, i.e.~a modified gravity signal is absent. This is a very
important case as it is a priori the most likely possible outcome of
the experiment. In this scenario, the parameter space outside the
confidence area of the posterior distribution can be ruled out. In the
case of no signal injection, we obtained an average error on the
determination of $a_0$ which is $\overline{\Delta a_0} = 0.055\cdot
10^{-10} \, \accunits$ for the current best estimate noise model and
$\overline{\Delta a_0} = 0.154\cdot{10^{-10}}\, \accunits$ for the
requirements noise. This rules out most values of $a_0$ except those
that are close to $0\, \accunits$. At the same time, there is a
complete uncertainty on $k$, which means that we will not be able to
draw any conclusions on this parameter in case of no signal detection.

In order to distinguish between signal detection and no signal
detection, we used the Bayesian approach to model selection. We
limited the choice to two models: one is the sum of noise and signal
(signal hypotheses), while the other consists of noise only (noise
hypothesis).  We computed the ratio of the probabilities for these two
hypotheses given the data and based on this number drew a conclusion
on which model is preferable.  We estimated the expectation for a
signal in the artificial data by calculating Bayes factors for $200$
different noise realisation for several points in the parameter space
listed in Table \ref{injection_points}.  On the basis of these
estimates, we were able to allocate areas in the parameter space where
the signal hypothesis could be strongly prioritised over the noise
hypothesis and areas where even in presence of a signal a confident
statement on its detection cannot be made. Notice that for a single
fly-by the experiment will provide us only with a single measured
dataset and a single deduced Bayes factor.  The estimates of the Bayes
factors for the artificial data gives a way to compare the single
Bayes factor estimated from the real data to the expected values and
judge the outcome of the experiment on the basis of this comparison.

Finally, we studied the data from one of the test campaigns for
\ac{LPF}. The importance of this study lies in the fact that in
reality the noise may have glitches and non-Gaussianities (see
Fig.~\ref{glitch_twoplots}).  When applied to this data, our Bayesian
model selection can prefer the signal hypothesis over the noise
hypothesis because neither of them describes the data with the glitch
correctly. In order to adequately address the problem of glitches, a
separate model to be fed to the Bayesian hypothesis selection approach
must be developed.

In our analysis we investigated the influence of the parameters $k$
and $a_0$ on the template, but we kept the interpolating function
fixed. As the interpolating function is heuristically designed on the
basis of astrophysical observations, it is not a smoothly varying
parameter but a point model. In a future work, we would like to apply
the data analysis framework we built to study a generalised,
phenomenological model of the interpolating function that uses a
finite set of parameters. This would allow us to assess different
theories that have \ac{MOND} as their non-relativistic
limit. Ultimately, the more general goal is to consider other theories
that yield a phenomenology detectable with \ac{LPF} and to be able to
perform a model selection among different models of gravity.

The significant issue left out of the scope of this paper is the
influence of the mission design and the mission time-line on the
experiment.  We leave it to future work to study the influence of the
accuracy of the acceleration recovery from the measurement of the
displacement on the parameter estimation.  Finally, the question of
how much data before and after the \ac{SP} fly-by needs to be gathered
to perform an accurate estimation of the acceleration and to assess
the possible non-Gaussianities in the noise is also left for future
work.

\section{Acknowledgements}

The authors would like to thank B. Sathyaprakash, Christian Trenkel,
Stephen Kemble, Badri Krishnan, Reinhard Prix, Ali Mozaffari, Luc
Blanchet, Paul McNamara, Gilles Esposito-Far\`{e}se, and Philippe
Jetzer for useful discussions, and Carsten Aulbert and Henning
Fehrmann for their comprehensive help. The numerical calculations were
performed on the \textit{Datura} cluster at the Max Planck Institute
for Gravitational Physics in Golm and on the \textit{Atlas} Cluster at
the Max Planck Institute for Gravitational Physics in Hannover. The
code used to solve the non-relativistic \ac{teves} equations was
kindly provided by the Imperial College London. The authors would like
to especially thank Jo\~{a}o Magueijo and Neil Bevis for providing the
code. The authors acknowledge Astrium Satellite GmbH (ASD), Germany
for providing the OSTT data from the \ac{LPF} test campaign. FP
acknowledges support from STFC Grant No.~ST/L000342/1 and DFG grant
SFB/Transregio~7.

\appendix
\section{\ac{teves}}
\label{appendix_teves}

\ac{teves} was the first consistent relativistic theory of gravity
reducing to \ac{MOND} in the non-relativistic limit. It is built upon
a nondynamical gravitational scalar field $\sigma$ and three dynamical
gravitational fields, namely, the Einstein metric tensor
$g_{\alpha\beta}$, a timelike $4$-vector field $\mathfrak{U}^{\beta}$,
and a scalar field $\phi$. Accordingly, it was dubbed \acl{teves}
theory. The physical metric may be obtained from the dynamical fields
via the relation $\tilde{g}_{\alpha\beta} = e^{-2\phi}g_{\alpha\beta}
- 2\mathfrak{U}_{\alpha}\mathfrak{U}_{\beta}\sinh(2\phi)$, where and
$\mathfrak{U}_{\alpha} = g_{\alpha\beta}\mathfrak{U}^{\beta}$.

Within this theory, the total action takes the form
\begin{equation}
  S = S_g + S_v + S_s + S_m\,,
\end{equation}
where $S_g$ is the Einstein-Hilbert action for the metric tensor,
$S_v$ is the action governing the timelike vector field, $S_s$ is the
action for the dynamical and the non-dynamical scalar fields, and
$S_m$ is the action for the matter fields. The equation for the
dynamical gravitational scalar field may be derived from
\begin{align}
  S_s & = -\frac{1}{2}\int [\sigma^2 h^{\alpha\beta} \phi_{,\alpha}
  \phi_{,\beta} + \frac{1}{2} G l^{-2} \sigma^4 F(kG\sigma^2) ] \sqrt{-g}\de^4 x\,,\nonumber\\
  \label{scalar_action}
\end{align}
where $g = \det(g_{\alpha \beta})$, $h^{\alpha \beta} \equiv
g^{\alpha\beta} - \mathfrak{U}^{\alpha} \mathfrak{U}^{\beta}$, $G$ is
the gravitational constant, $k$ is a dimensionless constant, $l$ is a
constant length, and $F$ is a free dimensionless function. Varying
$S_s$ with respect to the two scalar fields and using the equation for
$\sigma$ yields~\cite{TeVeS}
\begin{equation}
  \begin{split}
    & [\mu(kl^2h^{\mu\nu} \phi_{,\mu}\phi_{,\nu}) h^{\alpha\beta}\phi_{,\alpha}]_{;\beta} = \\
    & kG[g^{\alpha\beta} + (1 + e^{-4\phi})\mathfrak{U}^{\alpha}
    \mathfrak{U}^{\beta}] \tilde{T}_{\alpha\beta}\,,
    \label{phi_eom}
  \end{split}
\end{equation}
where $\tilde{T}_{\alpha\beta}$ is the physical energy-momentum
tensor, i.e. built upon the physical metric $\tilde{g}_{\alpha\beta}$,
and the function $\mu(y)$ obeys
\begin{equation}
  -\mu F(\mu) - \frac{1}{2} \mu^2 \frac{\de F(\mu)}{\de \mu} = y.
\end{equation}

\section{Newtonian Stress Tensor}
\label{appendix_Newtonian_ST}
\newcommand{\dsun}[0]{d_{\text{s}}} 
\newcommand{\dearth}[0]{d_{\text{e}}} 
\newcommand{\Msun}[0]{M_{\text{s}}} 
\newcommand{\Mearth}[0]{M_{\text{e}}} 
\newcommand{\rsunearth}[0]{r_{\text{se}}} 
\newcommand{\evec}[0]{\hat{\vc{e}}}

The expression of the Newtonian potential $\Phi_\text{N}$ for the Sun
-- Earth two-body system is
\begin{equation}
  \Phi_\text{N} = 
  -G \left[ \Mearth\frac{r_0-\dearth}{\dearth r_0} +
    \Msun \frac{\rsunearth - r_0 -\dsun}{\dsun (\rsunearth - r_0)}\right],
\end{equation}
where $G$ is Newton's gravitational constant, $\Msun$ ($M_e$) is the
mass of the Sun (Earth), $\rsunearth$ is the Sun -- Earth separation,
\begin{equation}
  r_0 = \frac{\rsunearth\sqrt{\Mearth/\Msun}}{\sqrt{\Mearth/\Msun}+1}  = 
  \frac{\rsunearth}{\sqrt{\Msun/\Mearth}+1}
\end{equation}
is the distance from the Earth to the \ac{SP}, and $\dsun$ ($\dearth$)
is the distance from the point where the potential is calculated to
the Sun (Earth) respectively, i.e.,
\begin{align}
  \dearth &= \sqrt{(x_1+r_0)^2+x_2^2+x_3^2},\\
  \dsun &= \sqrt{((\rsunearth - r_0)-x_1)^2 + x_2^2 + x_3^2}\,.
\end{align}
The gradient of the Newtonian potential is therefore
\begin{equation}
  \begin{split}
    \frac{\partial \Phi_\text{N}}{\partial x_i} &  = \frac{G\Mearth[x_i - r_0  c_i]}{\dearth^3}  \\
    + & \frac{G\Msun[x_i - (\rsunearth-r_0) c_i]}{\dsun^3}\,,
    \label{Newtonian_potential_gradient}
  \end{split}
\end{equation}
where $\evec_{x_i} (i=1..3)$ is the orthonormal unit vectors set of
the reference system and $ c_i = \evec_{x_1} \cdot \evec_{x_i}$. The
Newtonian stress tensor reads
\begin{equation}
  \begin{split}
    \label{Newtonian_st}
    \frac{\partial^2 \Phi_\text{N}}{\partial x_i^2} =& G \Mearth\left\{\frac{1}{\dearth^3} - \frac{3[x_i+r_0 c_i]^2}{\dearth^5}\right\} \\
    +& G \Msun \left\{\frac{1}{\dsun^3} - \frac{3[x_i - (\rsunearth-r_0) c_i]^2}{\dsun^5}\right\} \\
    \frac{\partial^2 \Phi_\text{N}}{\partial x_i \partial x_j} \Bigr\vert_{i\neq j} =& -\frac{3G\Mearth}{\dearth^5} [x_i + r_0  c_i][x_j + r_0  c_j] \\
    - \frac{3G\Msun}{\dsun^5} &[x_i - (\rsunearth - r_0) c_i][x_j -
    (\rsunearth - r_0) c_j]\,.
  \end{split}
\end{equation}

\bibliographystyle{apsrev4-1-noeprint.bst}
\bibliography{library}
\end{document}